\documentclass[12pt, titlepage]{article}
\usepackage{palatino}
\usepackage[affil-it]{authblk}
\usepackage[left=1in, right=1in, top=1in]{geometry}
\usepackage{hyperref}
\usepackage{natbib}
\usepackage{fancyhdr}
\usepackage{setspace} 
\usepackage{lastpage}
\usepackage{graphicx}	
\usepackage{amsmath,mathtools}	
\usepackage{amssymb,dsfont,bbm}	
\usepackage[ruled,vlined]{algorithm}  
\usepackage{xcolor}  

\DeclareMathAlphabet\mathbfcal{OMS}{cmsy}{b}{n} 
\DeclareMathOperator*{\argmin}{argmin}

\newcommand{\deriv}{\ensuremath{{\rm d}}}  

\newcommand{\Normal}[2]{\ensuremath{\mathcal{N}\left[{#1}, {#2} \right]}} 

\newcommand{\mean}[1]{\ensuremath{\mathbb{E}\left[{#1}\right]}}
\newcommand{\meanwrt}[2]{\ensuremath{\mathbb{E}_{{#2}}\left[{#1}\right]}}

\newcommand{\indicator}[1]{\ensuremath{\mathds{1}\left[{#1}\right]}}

\newcommand{\params}{\ensuremath{\boldsymbol\Theta}}
\newcommand{\data}{\ensuremath{\mathbf{D}}}

\newcommand{\likelihood}{\ensuremath{\mathcal{L}}}
\newcommand{\prior}{\ensuremath{\pi}}
\newcommand{\posterior}{\ensuremath{\mathcal{P}}}
\newcommand{\proposal}{\ensuremath{\mathcal{Q}}}
\newcommand{\evidence}{\ensuremath{\mathcal{Z}}}
\newcommand{\bayesfactor}{\ensuremath{\mathcal{R}}}
\newcommand{\credible}{\ensuremath{\mathcal{C}}}

\newcommand{\cov}{\ensuremath{\mathbf{C}}}
\newcommand{\meanvec}{\ensuremath{\boldsymbol{\mu}}}

\title{A Conceptual Introduction to Markov Chain Monte Carlo Methods}

\author[]{Joshua S. Speagle}

\affil[]{Center for Astrophysics\,\textbar\,Harvard \& Smithsonian, 60 Garden St., Cambridge, MA 02138, USA}

\date{jspeagle@cfa.harvard.edu}

\begin{document}

\maketitle

\begin{abstract}
Markov Chain Monte Carlo (MCMC) methods have become a cornerstone of
many modern scientific analyses by providing a straightforward
approach to numerically estimate uncertainties in the parameters of a model
using a sequence of random samples.
This article provides a basic introduction to MCMC methods
by establishing a strong conceptual understanding of
\textit{what} problems MCMC methods are trying to solve,
\textit{why} we want to use them, and
\textit{how} they work in theory and in practice.
To develop these concepts, I outline the foundations of Bayesian inference,
discuss how posterior distributions are used in practice,
explore basic approaches to estimate posterior-based quantities, and
derive their link to Monte Carlo sampling and MCMC.
Using a simple toy problem, I then demonstrate how these concepts
can be used to understand the benefits and drawbacks of various MCMC approaches.
Exercises designed to highlight various concepts are also included throughout the
article.
\end{abstract}



\section{Introduction} \label{sec:intro}

Scientific analyses generally rest on making inferences about
underlying physical models from various sources of observational data.
Over the last few decades, the quality and quantity of these
data have increased substantially as they become faster and cheaper to 
collect and store. At the same time, the same technology that has
made it possible to collect vast amounts of data has also led to
a substantial increase in the computational power and resources
available to analyze them. 

Together, these changes have made it possible to explore
increasingly complex models using methods that
can exploit these computational resources. This has led to
a dramatic rise in the number of published works that rely on
\textbf{Monte Carlo} methods, which use
a combination of numerical simulation and random number generation
to explore these models.

One particularly popular subset of Monte Carlo methods is known as
\textbf{Markov Chain Monte Carlo (MCMC)}. MCMC methods are appealing
because they provide a straightforward, intuitive way to both simulate values
from an unknown distribution and use those simulated values
to perform subsequent analyses.
This allows them to be applicable in a wide variety of domains.

Owing to its widespread use, various overviews of MCMC methods are
common both in peer-reviewed and non-peer-reviewed sources.
In general, these tend to
fall into two groups: articles focused on various statistical underpinnings
of MCMC methods and articles focused on implementation and practical usage.
Readers interested in reading more details on either topic are encouraged
to see \citet{brooks+11} and \citet{hoggforemanmackey18_alt} along with
associated references therein.

This article instead provides an overview of MCMC methods focused
instead on building up a strong \textit{conceptual understanding} of
the what, why, and how of MCMC based on statistical intuition.
In particular, it tries to systematically answer the following questions:
\begin{enumerate}
    \item \textit{What} problems are MCMC methods trying to solve?
    \item \textit{Why} are we interested in using them?
    \item \textit{How} do they work in theory and in practice?
\end{enumerate}

When answering these questions, this article generally assumes that
the reader is somewhat familiar with
the basics of Bayesian inference in theory (e.g., the role of priors)
and in practice (e.g., deriving posteriors), 
basic statistics (e.g., expectation values), 
and basic numerical methods (e.g., Riemann sums).
No advanced statistical knowledge is required. For more details
on these topics, please see \citet{gelman+13} and \citet{blitzsteinhwang14}
along with associated references therein.

The outline of the article is as follows. 
In \S\ref{sec:bayes}, I provide a brief review of Bayesian inference
and posterior distributions. 
In \S\ref{sec:what}, I discuss what posteriors are used for in practice,
focusing on integration and marginalization.
In \S\ref{sec:grid}, I outline a basic scheme to approximate
these posterior integrals using discrete grids.
In \S\ref{sec:montecarlo}, I illustrate how Monte Carlo methods
emerge as a natural extension of grid-based approaches.
In \S\ref{sec:mcmc}, I discuss how MCMC methods fit within
the broader scope of possible approaches and their benefits and drawbacks.
In \S\ref{sec:sampling}, I explore the general challenges MCMC methods face.
In \S\ref{sec:example}, I examine how these concepts come together
in practice using a simple example.
I conclude in \S\ref{sec:conc}.

\section{Bayesian Inference} \label{sec:bayes}

In many scientific applications, we have access to some \textbf{data} $\data$
that we want to use to make inferences about the world around us.
Most often, we want to interpret these data in light of an underlying
\textbf{model} $M$ that can make predictions about the data we expect
to see as a function of some \textbf{parameters} $\params_M$ of that
particular model.

We can combine these pieces together to estimate the
\textbf{probability} $P(\data|\params_M, M)$ 
that we would actually see that data $\data$ we have
collected \textit{conditioned on} (i.e. assuming)
a specific choice of parameters $\params_M$ from our model $M$.
In other words, assuming our model $M$ is right and
the parameters $\params_M$ describe the data, what is the
\textbf{likelihood} $P(\data|\params_M, M)$ of the parameters $\params_M$
based on the observed data $\data$? Assuming different values
of $\params_M$ will give different likelihoods, telling us
which parameter choices appear to best describe the data we
observe.

In Bayesian inference, we are interested in
inferring the flipped quantity, $P(\params_M|\data, M)$.
This describes the probability that the 
underlying \textit{parameters} are actually $\params_M$ 
given our data $\data$ and assuming a particular model $M$.
By using factoring of probability, we can relate this new
probability $P(\params_M|\data, M)$ to the likelihood
$P(\data|\params_M, M)$ described above as
\begin{equation}
    P(\params_M | \data, M) P(\data | M) 
    = P(\params_M, \data | M)
    = P(\data | \params_M, M) P(\params_M | M)
\end{equation}
where $P(\params_M, \data | M)$ represents the \textit{joint}
probability of having an underlying set of parameters $\params_M$
that describe the data and observing the particular set of data $\data$
we have already collected.

Rearranging this equality into a more convenient form 
gives us \textbf{Bayes' Theorem}:
\begin{equation}
    P(\params_M | \data, M) 
    = \frac{P(\data | \params_M, M) P(\params_M | M)}{P(\data | M)}
\end{equation}
This equation now describes exactly how our two probabilities relate to each other.

$P(\params_M | M)$ is often referred to as the \textbf{prior}. This describes
the probability of having a particular set of values $\params_M$
for our given model $M$ \textit{before conditioning on our data}.
Because this is independent of the data, this term
is often interpreted as representing 
our ``prior beliefs'' about what $\params_M$ \textit{should}
be based on previous measurements, physical concerns, 
and other known factors. In practice, this has the effect of essentially
``augmenting'' the data with other information.

The denominator
\begin{equation}
    P(\data | M) = \int P(\data | \params_M, M) P(\params_M | M) \deriv \params_M
\end{equation}
is known as the \textbf{evidence} or marginal likelihood
for our model $M$ \textbf{marginalized} (i.e. integrated)
over all possible parameter values $\params_M$.
This broadly tries to quantify
how well our model $M$ explains the data $\data$ after averaging over
all possible values $\params_M$ of the true underlying parameters.
In other words, if the observations predicted by our model look similar
to the data $\data$, then $M$ is a good model.
Models where this is true more often also tend to be favored 
over models that give excellent agreement occasionally but 
disagree most of the time. Since in most instances we take
$\data$ as a given, this often ends up being a constant.

\begin{figure}
\begin{center}
\includegraphics[width=\textwidth]{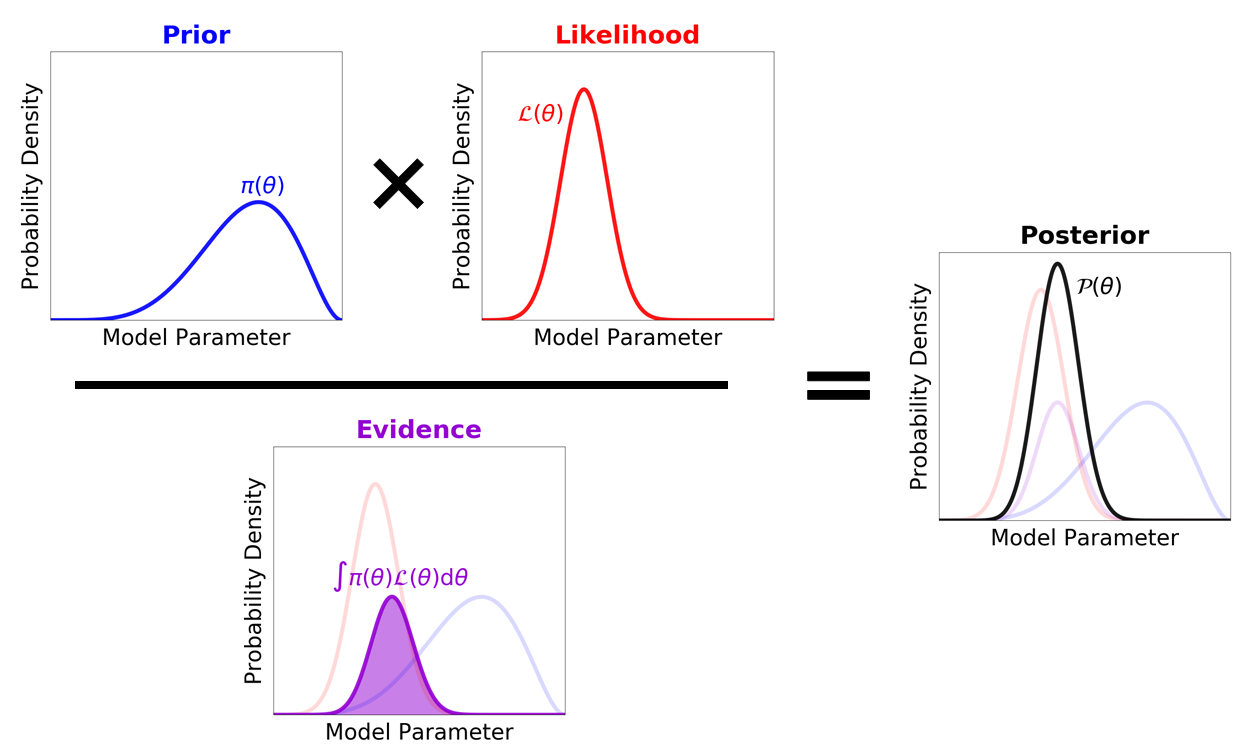}
\end{center}
\caption{An illustration of Bayes' Theorem. The posterior probability $\posterior(\params)$
(black) of our model parameters $\params$ is based on a combination of our
prior beliefs $\prior(\params)$ (blue) and the likelihood $\likelihood(\params)$ 
(red), normalized by the overall evidence 
$\evidence = \int \prior(\params) \likelihood(\params) \deriv \params$ (purple)
for our particular model. See \S\ref{sec:bayes} for additional details.
}\label{fig:bayes}
\end{figure}

Finally, $P(\params_M | \data, M)$ represents our \textbf{posterior}.
This quantifies our belief in $\params_M$ after combining our
prior intuition $P(\params_M|M)$ with
current observations $P(\data|\params_M,M)$ 
and normalizing by the overall evidence $P(\data|M)$.
The posterior will be some compromise between the prior and 
the likelihood, with the exact combination depending
on the strength and properties of the prior and the quality
of the data used to derive the likelihood.
A schematic illustration is shown in 
{\color{red} \autoref{fig:bayes}}.

Throughout the rest of the paper I will write 
these four terms (likelihood, prior, evidence, posterior)
using shorthand notation such that
\begin{equation}
    \posterior(\params) 
    \equiv \frac{\likelihood(\params)\prior(\params)}{
    \int \likelihood(\params)\prior(\params) \deriv \params}
    \equiv \frac{\likelihood(\params)\prior(\params)}{\evidence}
\end{equation}
where $\posterior(\params) \equiv P(\params_M | \data, M)$ is the posterior,
$\likelihood(\params) \equiv P(\data | \params_M, M)$ is the likelihood,
$\prior(\params) \equiv P(\params_M | M)$ is the prior, and the constant
$\evidence \equiv P(\data | M)$ is the evidence. I have suppressed the
model $M$ and data $\data$ notation
for convenience here since in most cases the data and model are considered
fixed, but will re-introduce them as necessary.

Before moving on, I would like to close by emphasizing that
\textit{the interpretation of any result is
only as good as the models and priors that underlie them}.
Trying to explore the implications of any particular model
using, for instance, some of the methods described in this article
is fundamentally a \textit{secondary concern} behind constructing 
a reasonable model with well-motivated priors in the first place. 
I strongly encourage readers to keep this idea in mind
throughout the remainder of this work.

\subsection*{Exercise: Noisy Mean} \label{exercise:bayes}

\subsubsection*{Setup}

Consider the case where we have temperature monitoring stations
located across a city. Each station $i$ takes a noisy measurement
$\hat{T}_i$ of the temperature on any given day with some
measurement noise $\sigma_i$. We will assume our measurements $\hat{T}_i$
follow a \textbf{Normal (i.e. Gaussian) distribution}
with mean $T$ and standard deviation $\sigma_i$ such that
\begin{equation*}
    \hat{T}_i \sim \Normal{T}{\sigma_i}
\end{equation*}
This translates into a probability of
\begin{equation*}
    P(\hat{T}_i|T,\sigma_i) \equiv \Normal{T}{\sigma_i}
    = \frac{1}{\sqrt{2\pi\sigma_i^2}}
    \exp\left[-\frac{1}{2}\frac{(\hat{T}_i - T)^2}{\sigma_i^2}\right]
\end{equation*}
for each observation and
\begin{equation*}
    P(\{ \hat{T}_i \}_{i=1}^{n} | T, \{ \sigma_i \}_{i=1}^{n})
    = \prod_{i=1}^{n} P(\hat{T}_i|T,\sigma_i)
\end{equation*}
for a collection of $n$ observations.

Let's assume we have five independent noisy measurements
of the temperature (in Celsius) from several monitoring stations
\begin{equation*}
    \hat{T}_1 = 26.3, \: \hat{T}_2 = 30.2, \: \hat{T}_3 = 29.4, 
    \hat{T}_4 = 30.1, \: \hat{T}_5 = 29.8
\end{equation*}
with corresponding uncertainties
\begin{equation*}
    \sigma_1 = 1.7, \: \sigma_2 = 1.8, \: \sigma_3 = 1.2, 
    \sigma_4 = 0.5, \: \sigma_5 = 1.3
\end{equation*}

Looking at historical data, we find that the typical underlying
temperature $T$ during similar days is roughly Normally-distributed
with a mean $T_{\rm prior}=25$ and variation $\sigma_{\rm prior} = 1.5$:
\begin{equation*}
    T \sim \Normal{T_{\rm prior}=25}{\sigma_{\rm prior}=1.5}
\end{equation*}

\subsubsection*{Problem}

Using these assumptions, compute:
\begin{enumerate}
    \item the prior $\prior(T)$,
    \item the likelihood $\likelihood(T)$, and
    \item the posterior $\posterior(T)$
\end{enumerate}
given our observed data $\{ \hat{T}_i \}$ and errors $\{ \sigma_i \}$
over a range of temperatures $T$. How do the three terms differ?
Does the prior look like a good assumption? Why or why not?

\section{What are Posteriors Good For?} \label{sec:what}

Above, I described how Bayes' Theorem is able to combine
our prior beliefs and the observed data into a new posterior estimate 
$\posterior(\params) \propto \likelihood(\params) \prior(\params)$.
This, however, is only half of the problem. Once we have
the posterior, we need to then \textit{use} it to
make inferences about the world around us. 
In general, the ways in which we want to use posteriors
fall into a few broad categories:
\begin{enumerate}
    \item \textbf{Making educated guesses}:
    make a reasonable guess 
    at what the underlying model parameters are.
    \item \textbf{Quantifying uncertainty}:
    provide constraints on 
    the range of possible model parameter values.
    \item \textbf{Generating predictions}:
    marginalize over uncertainties in the underlying model parameters 
    to predict observables or other variables that depend on the
    model parameters.
    \item \textbf{Comparing models}:
    use the evidences from different models
    to determine which models are more favorable.
\end{enumerate}

In order to accomplish these goals,
we are often more interested in trying to use the
posterior to estimate various constraints on the parameters
$\params$ themselves or other quantities $f(\params)$ that
might be based on them. This often depends on marginalizing
over the uncertainties characterized by our posterior (via the likelihood
and prior). The evidence $\evidence$, for instance, is
again just the integral of the likelihood and the prior 
over all possible parameters:
\begin{equation}
    \evidence 
    = \int \likelihood(\params) \prior(\params) \deriv \params
    \equiv \int \tilde{\posterior}(\params) \deriv \params
\end{equation}
where $\tilde{\posterior}(\params) \equiv \likelihood(\params) \prior(\params)$
is the \textit{unnormalized} posterior.

Likewise, if we are investigating the behavior of a subset of
``interesting'' parameters $\params_{\rm int}$ from 
$\params = \{ \params_{\rm int}, \params_{\rm nuis} \}$,
we want to marginalize over the behavior of the remaining 
``nuisance'' parameters $\params_{\rm nuis}$ to see how
they can impact $\params_{\rm int}$. This process is pretty
straightforward if the entire posterior over $\params$ is known:
\begin{equation}
    \posterior(\params_{\rm int})
    = \int \posterior(\params_{\rm int}, \params_{\rm nuis}) \, \deriv \params_{\rm nuis}
    = \int \posterior(\params) \deriv \params_{\rm nuis}
\end{equation}

Other quantities can generally be derived from the \textbf{expectation value}
of various parameter-dependent functions $f(\params)$ with respect to the posterior:
\begin{equation}
    \meanwrt{f(\params)}{\posterior} 
    \equiv \frac{\int f(\params) \posterior(\params) \deriv \params}
    {\int \posterior(\params) \deriv \params} 
    = \frac{\int f(\params) \tilde{\posterior}(\params) \deriv \params}
    {\int \tilde{\posterior}(\params) \deriv \params} 
    = \int f(\params) \posterior(\params) \deriv \params
\end{equation}
since $\int \posterior(\params) \deriv \params = 1$ by definition 
and $\tilde{\posterior}(\params) \propto \posterior(\params)$.
This represents a weighted average of $f(\params)$,
where at each value $\params$ we weight the resulting
$f(\params)$ based on to the 
chance we believe that value is correct.

Taken together, we see that in almost all cases \textit{we are
more interested in computing integrals over the posterior
rather than knowing the posterior itself.}
To put this another way, the posterior is rarely ever useful on its
own; it mainly becomes useful by integrating over it.


This distinction between estimating expectations and other integrals
over the posterior versus estimating the posterior 
in-and-of-itself is a key element of Bayesian inference.
This distinction is hugely important when it comes
to actually performing inference in practice, since it
is often the case that we can get an excellent estimate of
$\meanwrt{f(\params)}{\posterior}$ even if we have an
extremely poor estimate of $\posterior(\params)$ 
or $\tilde{\posterior}(\params)$.

\begin{figure}
\begin{center}
\includegraphics[width=\textwidth]{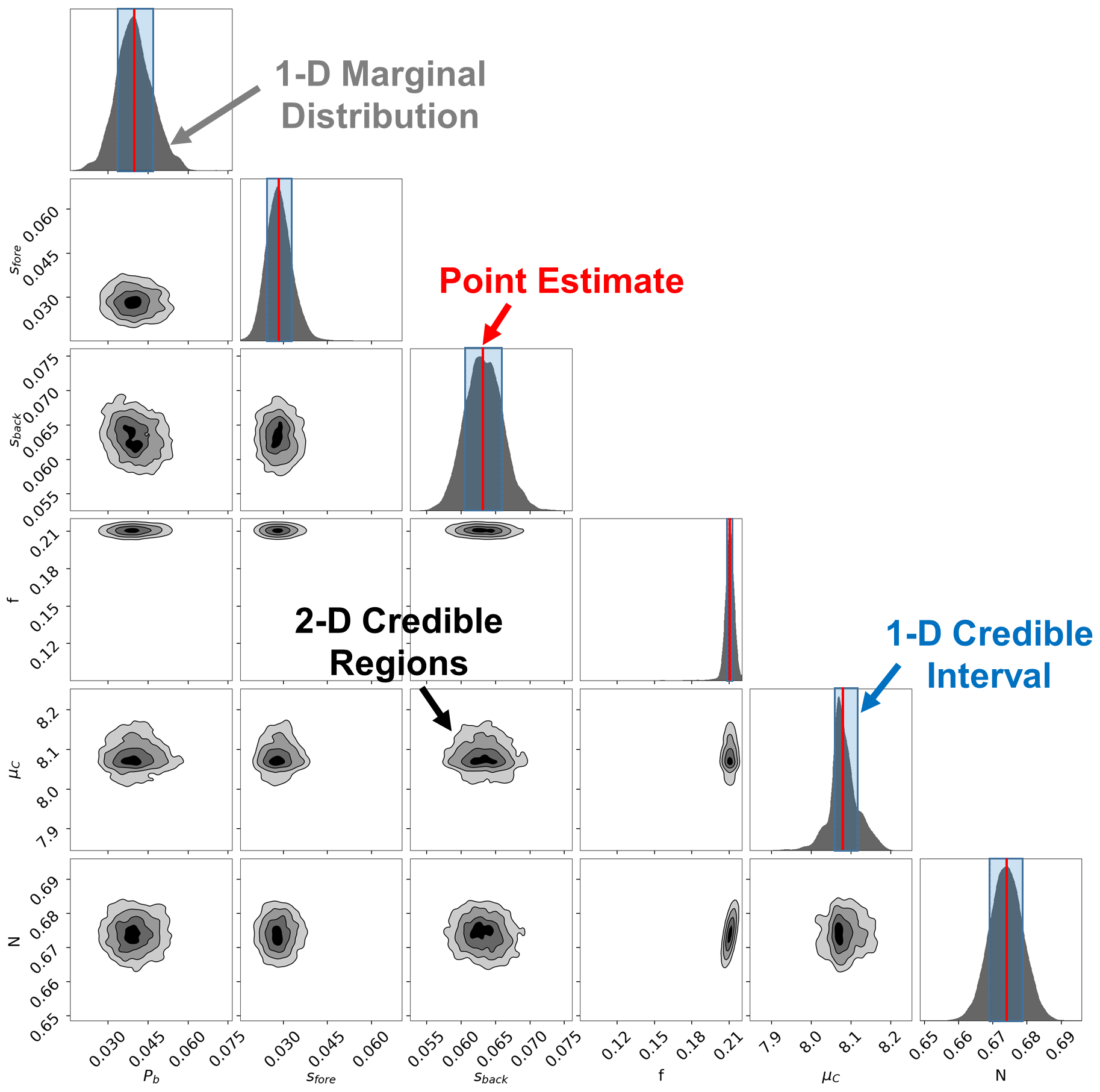}
\end{center}
\caption{A ``corner plot'' showing an example of how posteriors
are used in practice. Each of the top panels shows the
1-D marginalized posterior distribution for each parameter
(grey), along with associated median point estimates
(red) and 68\% credible intervals (blue). Each central
panel shows the 10\%, 40\%, 65\%, and 85\% credible regions
for each 2-D marginalized posterior distribution.
See \S\ref{sec:what} for additional details.
}\label{fig:corner}
\end{figure}

More details are provided below to further illustrate how
the particular categories described above translate into
particular integrals over the (unnormalized) posterior.
An example is shown in {\color{red} \autoref{fig:corner}}.

\subsection{Making Educated Guesses} \label{subsec:guess}

One of the core tenets of Bayesian inference is that
we don't know the true model $M_*$ or its true underlying parameters
$\params_*$ that characterize the data we observe: the model $M$
we have is almost always a simplification of what is actually going on.
If we assume that our current model $M$ is correct, however,
we can try to use our posterior $\posterior(\params)$
to propose a \textbf{point estimate} $\hat{\params}$
that we think is a pretty good guess for the true value $\params_*$.

What exactly counts as ``good''? This depends on exactly what we care
about. In general, we can quantify ``goodness'' by asking the opposite question:
how badly are we penalized if our estimate $\hat{\params} \neq \params_*$
is wrong? This is often encapsulated through the use of
a \textbf{loss function} $L(\hat{\params}|\params_*)$
that penalizes us when our point estimate $\hat{\params}$
differs from $\params_*$. An example of a common loss function
is $L(\hat{\params}|\params_*) = |\hat{\params} - \params_*|^2$
(i.e. squared loss), where an incorrect guess is penalized based on
the square of the magnitude of the separation between the
guess $\hat{\params}$ and the true value $\params_*$.

Unfortunately, we don't know what the actual value of
$\params_*$ is to evaluate the true loss. We can, however,
do the next best thing and compute the \textbf{expected loss}
averaged over all possible values of $\params_*$ based on
our posterior:
\begin{equation}
    L_\posterior(\hat{\params})
    \equiv \meanwrt{L(\hat{\params}|\params)}{\posterior}
    = \int L(\hat{\params}|\params) \posterior(\params) \deriv \params
\end{equation}
A reasonable choice for $\hat{\params}$ is then
the value that minimizes this expected loss in place of the
actual (unknown) loss:
\begin{equation}
    \hat{\params} 
    \equiv \argmin_{\params'} \left[L_\posterior(\params')\right]
\end{equation}
where $\argmin$ indicates the value (argument) of $\params'$
that minimizes the expected loss $L_\posterior(\params')$.

While this strategy can work for any arbitrary loss function, solving
for $\hat{\params}$ often requires using numerical methods and repeated
integration over $\posterior(\params)$. However, analytic solutions
do exist for particular loss functions. For example,
it is straightforward to show 
(and an insightful exercise for the interested reader) 
that the optimal point estimate $\hat{\params}$
under squared loss is simply the mean.

\subsection{Quantifying Uncertainty} \label{subsec: guess}

In many cases we are not just interested in computing
a prediction $\hat{\params}$ for $\params_*$, but also constraining
a region $\credible(\params)$ of possible values 
within which $\params_*$ might lie with some amount of certainty.
In other words, can we construct a region $\credible_X$
such that we believe there is an $X\%$ chance that it contains
$\params_*$?

There are many possible definitions
for this \textbf{credible region}. One common definition
is the region above some posterior threshold $\posterior_X$
where $X\%$ of the posterior is contained, i.e. where
\begin{equation}
    \int_{\params \,\in\, \credible_X} \posterior(\params) \deriv \params
    = \frac{X}{100}
\end{equation}
given
\begin{equation}
    \credible_X
    \equiv \left\{ \params : \posterior(\params) \geq \posterior_X \right\}
\end{equation}

In other words, we want to integrate our posterior over all $\params$
where the value $\posterior(\params) > \posterior_X$ is greater than some threshold
$\posterior_X$, where $\posterior_X$ is set so 
that this integral encompasses $X\%$ of the full posterior.
Common choices for $X$ include $68\%$ and $95\%$ (i.e.
``1-sigma'' and ``2-sigma'' credible intervals).

In the special case where our (marginalized) posterior is 1-D, \textbf{credible intervals} 
are often defined using \textbf{percentiles} rather than thresholds, where
the location $x_p$ of the $p$th percentile is defined as
\begin{equation}
    \int_{-\infty}^{x_p} \posterior(x) \deriv x = \frac{p}{100}
\end{equation}
We can use these to define a credible region $[x_{\rm low}, x_{\rm high}]$
containing $Y\%$ of the data by taking $x_{\rm low} = x_{(1-Y)/2}$ and
$x_{\rm high} = x_{(1+Y)/2}$. While this leads to asymmetric thresholds
and does not generalize to higher dimensions, it has the benefit of
always encompassing the median value $x_{50}$ and having equal tail
probabilities (i.e. $(1-Y)/2\%$ of the posterior on each side).

In general, when referring to ``credible intervals'' throughout the text
the percentile definition should be assumed unless explicitly stated otherwise.

\subsection{Making Predictions} \label{subsec:pred}

In addition to trying to estimate the underlying parameters of our
model, we often also want to make predictions of other observables or variables
that depend on our model parameters.
If we think we know the underlying true model parameters $\params_*$, then
this process is straightforward. Given that we only have access to the
posterior distribution $\posterior(\params)$ over possible
values $\params_*$ could take, however, to predict what will happen we
will need to marginalize over this uncertainty.

We can quantify this intuition using the \textbf{posterior predictive}
$P(\tilde{\data}|\data)$, which represents the probability
of seeing some new data $\tilde{\data}$ based on our existing data $\data$:
\begin{equation}
    P(\tilde{\data}|\data) 
    \equiv \int P(\tilde{\data}|\params) P(\params|\data) \deriv \params
    \equiv \int \tilde{\likelihood}(\params) \posterior(\params) \deriv \params
    = \meanwrt{\tilde{\likelihood}(\params)}{\posterior}
\end{equation}
In other words, for hypothetical data $\tilde{\data}$, 
we want to compute the expected value of the likelihood
$\tilde{\likelihood}(\params)$ over all possible
values of $\params$ based on the current posterior $\posterior(\params)$.

\subsection{Comparing Models} \label{subsec:evid}

One final point of interest in many Bayesian analyses is
trying to investigate whether the data particularly favors
any of the model(s) we are assuming in our analysis.
Our choice of priors or the particular way we parameterize the
data can lead to substantial differences in the way we
might want to interpret our results.

We can compare two models by computing the \textbf{Bayes factor}:
\begin{equation}
    \bayesfactor^{1}_{2}
    \equiv \frac{P(M_{\rm 1}|\data)}{P(M_{\rm 2}|\data)}
    = \frac{P(\data|M_{\rm 1})P(M_{\rm 1})}{P(\data|M_{\rm 2})P(M_{\rm 2})}
    \equiv \frac{\evidence_{\rm 1}}{\evidence_{\rm 2}} 
    \frac{\prior_{\rm 1}}{\prior_{\rm 2}}
\end{equation}
where $\evidence_M$ is again the evidence for model $M$ 
and $\prior_M$ is our prior belief that $M$ 
is correct relative to the competing model.
Taken together, the Bayes factor $\bayesfactor$ tells us
how much a particular model is favored over another given the
observed data, marginalizing over all possible values of the underlying
model parameters $\params_M$, and our previous relative confidence in the model.

Again, note that computing $\evidence_M$ requires computing the
integral $\int \tilde{\posterior}(\params) \deriv \params$ of the
unnormalized posterior $\tilde{\posterior}(\params)$ over $\params$. 
Combined with the other examples outlined in this section,
it is clear that many common use cases in Bayesian analysis rely on
computing integrals over the (possibly unnormalized) posterior.

\subsection*{Exercise: Noisy Mean Revisited} \label{exercise:practice}

\subsubsection*{Setup}

Let's return to our temperature posterior $\posterior(T)$ from
\S\ref{exercise:bayes}. We want to use this result to derive interesting
estimates and constraints on the possible underlying temperature $T$.

\subsubsection*{Point Estimates}

The \textbf{mean} can be defined as the
point estimate $\hat{\params}$ that minimizes the expected loss
$L_{\posterior}(\hat{\params})$ under \textbf{squared loss}:
\begin{equation*}
    L_{\rm mean}(\hat{\params}|\params_*) = |\hat{\params} - \params_*|^2
\end{equation*}
The \textbf{median} can be defined as the point estimate that minimizes
$L_{\posterior}(\hat{\params})$ under \textbf{absolute loss}:
\begin{equation*}
    L_{\rm med}(\hat{\params}|\params_*) = |\hat{\params} - \params_*|
\end{equation*}
And the \textbf{mode} can be defined as the point estimate that
minimizes $L_{\posterior}(\hat{\params})$ under \textbf{``catastrophic'' loss}:
\begin{equation*}
    L_{\rm mode}(\hat{\params}|\params_*) = -\delta(|\hat{\params}-\params_*|)
\end{equation*}
where $\delta(\cdot)$ is the \textbf{Dirac delta function} defined such that
\begin{equation*}
    \int f(x) \delta(x-a) \deriv x = f(a)
\end{equation*}

Given these expressions for the mean, median,
and mode, estimate the corresponding temperature point estimate
$T_{\rm mean}$, $T_{\rm med}$, and $T_{\rm mode}$ from our corresponding
posterior. Feel free to experiment with various analytic and numerical
methods to perform these calculations.

We might expect that the historical data we used for our priors
might not hold as well today if there have been some long-term changes
in the average temperature.
For instance, we expect that the average temperature has increased
over time, and so we might not want to penalize hotter temperatures
$T \geq T_{\rm prior}$ as much as cooler ones $T < T_{\rm prior}$.
We can encode this information in an asymmetric loss function such as
\begin{equation*}
    L(\hat{T}|T_*) = 
    \begin{cases}
    |\hat{T} - T_*|^3 & T < T_{\rm prior} \\
    |\hat{T} - T_*| & T \geq T_{\rm prior}
    \end{cases}
\end{equation*}
What is the optimal point estimate $T_{\rm asym}$ that minimizes
the expected loss in this case?

\subsubsection*{Credible Intervals}

Next, let's try to quantify the uncertainty. Given the posterior
$\posterior(T)$, compute the 50\%, 80\%, and 95\% credible intervals
using posterior thresholds $\posterior_X$. Next, compute these
credible intervals using percentiles. Are there are differences between
the credible intervals computed from the two methods? Why or why not?

\subsubsection*{Posterior Predictive}

To propagate our uncertainties into the next observations,
compute the posterior predictive $P(\hat{T}_6|\{ \hat{T}_1, \dots, \hat{T}_5 \})$ 
over a range of possible temperature measurements $\hat{T}_6$ 
for the next observations given the previous five
$\{ \hat{T}_1, \dots, \hat{T}_5 \}$ assuming
an uncertainty of $\sigma_6=0$, $\sigma_6=0.5$, and $\sigma_6=2$.

\subsubsection*{Model Comparison}

Finally, we want to investigate whether our prior appears to be a good
assumption. Using numerical methods, compute the
evidence $\evidence$ for our default prior with mean $T_{\rm prior} = 25$
and standard deviation $\sigma_{\rm prior} = 1.5$. Then compare this to the
evidence estimated based on an alternative prior
where we assume the temperature has risen by roughly five
degrees with mean $T_{\rm prior} = 30$ but with a corresponding larger uncertainty
$\sigma_{\rm prior} = 3$. Is one model particularly favored over the other?

\section{Approximating Posterior Integrals with Grids} \label{sec:grid}

I now want to investigate methods for estimating posterior integrals.
While in some cases (e.g., conjugate priors) 
these can be computed analytically, this is not true in general.
To properly estimate quantities such as those outlined in
\S\ref{sec:what} therefore requires the use of numerical methods
(highlighted in the previous exercises).

\begin{figure}
\begin{center}
\includegraphics[width=\textwidth]{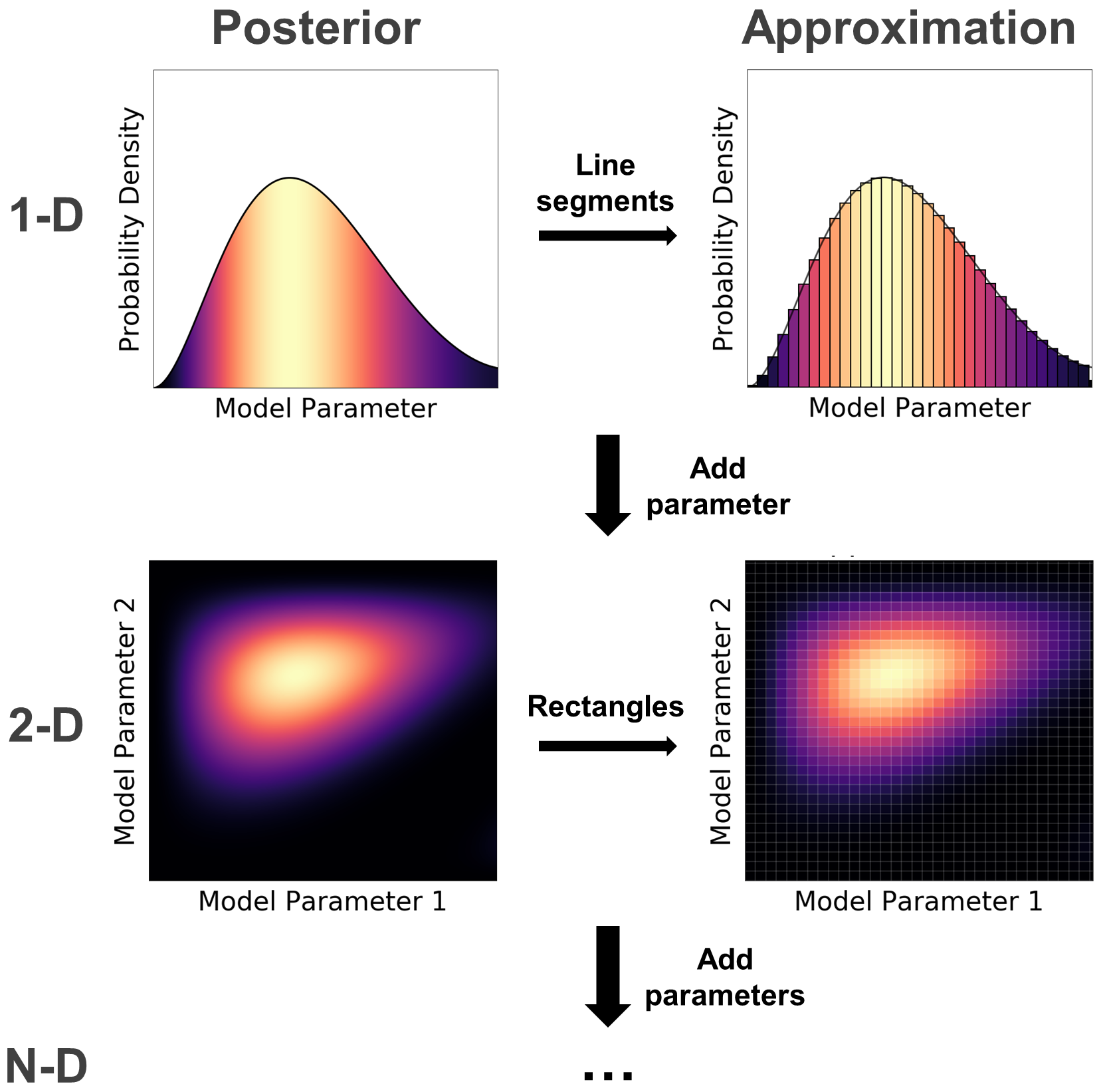}
\end{center}
\caption{An illustration of how to approximate posterior integrals
using a discrete grid of points. We break up the posterior
into contiguous regions defined by a position $\params_i$ (e.g.,
an endpoint or midpoint) with corresponding posterior density
$\posterior(\params_i)$ and volume $\Delta \params_i$
over a grid with $i=1,\dots,n$ elements. Our integral can then
be approximated by adding up each of these regions proportional
to the posterior mass $\posterior(\params_i) \times \Delta \params_i$
contained within it. In 1-D (top), these volume elements $\Delta\params_i$
correspond to line segments while in 2-D (middle), these 
correspond to rectangles. This can be generalized to higher dimensions
(bottom), where we instead used N-D cuboids.
See \S\ref{sec:grid} for additional details.
}\label{fig:riemann}
\end{figure}

To start, I will first focus on the case where our integral over
$\params$ is 1-D. In that case, we can approximate it using standard
numerical techniques such as a \textbf{Riemann sum} over a \textbf{discrete grid}
of points:
\begin{equation}
    \meanwrt{f(\params)}{\posterior} 
    = \int f(\params) \posterior(\params) \deriv \params
    \approx \sum_{i=1}^{n} f(\params_i) 
    \posterior(\params_i) \Delta \params_i
\end{equation}
where
\begin{equation}
    \Delta \params_i = \params_{j+1} - \params_{j}
\end{equation}
is simply the spacing between the set of $j=1,\dots,n+1$ points
on the underlying grid and
\begin{equation}
    \params_i = \frac{\params_{j+1} + \params_{j}}{2}
\end{equation}
is just defined to be the mid-point between $\params_j$ and
$\params_{j+1}$.\footnote{Choosing $\params_i$ to be one of the end-points 
gives consistent behavior (see \S\ref{subsec:consistent}) as 
the number of grid points $n \rightarrow \infty$ but generally leads
to larger biases for finite $n$.} As shown in 
{\color{red} \autoref{fig:riemann}}, 
this approach is akin to trying to approximate the
integral using a discrete set of $n$ rectangles with heights of
$f(\params_i) \posterior(\params_i)$ and widths of $\Delta \params_i$.

This idea can be generalized to higher dimensions. In that case,
instead of breaking up the integral into $n$ 1-D segments, 
we instead can decompose it into a set of $n$ N-D cuboids.
The contribution of each of these pieces is then proportional
to the product of the ``height'' $f(\params_i) \posterior(\params_i)$
and the \textit{volume}
\begin{equation}
    \Delta \params_i = \prod_{j=1}^{d} \Delta \Theta_{i,j}
\end{equation}
where $\Delta \Theta_{i,j}$ is the width of the $i$th cuboid in the $j$th dimension.
See {\color{red} \autoref{fig:riemann}}
for a visual representation of this procedure.

Substituting $\posterior(\params) = \tilde{\posterior}(\params)/\evidence$
into the expectation value and replacing any integrals 
with their grid-based approximations then gives:
\begin{equation} \label{eqn:exp_grid}
    \meanwrt{f(\params)}{\posterior}
    = \frac{\int f(\params) {\posterior}(\params) \deriv \params}
    {\int {\posterior}(\params) \deriv \params}
    = \frac{\int f(\params) \tilde{\posterior}(\params) \deriv \params}
    {\int \tilde{\posterior}(\params) \deriv \params}
    \approx \frac{\sum_{i=1}^{n} f(\params_i) 
    \tilde{\posterior}(\params_i) \Delta \params_i}
    {\sum_{i=1}^{n}\tilde{\posterior}(\params_i) \Delta \params_i}
\end{equation}
Note the denominator is now an estimate for the evidence:
\begin{equation}
    \evidence 
    = \int \tilde{\posterior}(\params) \deriv \params
    \approx \sum_{i=1}^{n} \tilde{\posterior}(\params_i) \Delta \params_j
\end{equation}
This substitution of the unnormalized posterior
$\tilde{\posterior}(\params)$
for the posterior $\posterior(\params)$ is a crucial part of computing 
expectation values in practice since we can compute 
$\tilde{\posterior}(\params) = \likelihood(\params) \prior(\params)$
directly without knowing $\evidence$.

\subsection{The Curse of Dimensionality} \label{subsec:curse}

While this approach is straightforward, it has one immediate
and severe drawback: the total number of grid points increases
\textit{exponentially} as the number of dimensions increases. 
For example, assuming we have roughly $k \geq 2$
grid points in each dimensions, the total number of points $n$ in our
grid scales as
\begin{equation}
    n \sim \prod_{j=1}^{d} k = k^d
\end{equation}
This means that even in the absolute \textit{best} case where $k=2$,
we have $2^d$ scaling.

This awful scaling is often referred to as the \textbf{curse of dimensionality}.
This exponential dependence turns out to be a generic feature 
of high-dimensional distributions
(i.e. posteriors of models with larger numbers of parameters)
that I will return to later in \S\ref{sec:sampling}.

\subsection{Effective Sample Size} \label{subsec:ess}

Apart from this exponential scaling of dimensionality, there is
a more subtle drawback to using grids. Since we do not
know the shape of the distribution ahead of time, the contribution of
each portion of the grid (i.e. each N-D cuboid)
can be highly uneven depending on the structure of the grid. 
In other words, the effectiveness of this approach not
only depends on the \textit{number} of grid points $n$ but 
also \textit{where} they are allocated. 
If we do not specify our grid points well,
we can end up with many points located in regions where
$\tilde{\posterior}(\params)$ and/or $f(\params)\tilde{\posterior}(\params)$
is relatively small. This then implies that their respective sums 
will be dominated by a small number of points with 
much larger relative ``weights''. Ideally, we would
want to increase the resolution of the grid
in regions where the posterior is large and decrease it elsewhere
to mitigate this effect. 

Note that our use of the term ``weights'' in the preceding paragraph 
is quite deliberate. Looking back at our original approximation,
the form of equation \eqref{eqn:exp_grid} 
is quite similar to one which might be used to compute a 
\textbf{weighted sample mean} of $f(\params)$. 
In that case, where we have $n$ observations $\{ f_1, \dots, f_n \}$
with corresponding weights $\{ w_1, \dots, w_n \}$, the
weighted mean is simply:
\begin{equation} \label{eqn:wt_mean}
    \hat{f}_{\rm mean} \equiv \frac{\sum_{i=1}^{n} w_i f_i}
    {\sum_{i=1}^{n} w_i}
\end{equation}
Indeed, if we define
\begin{equation}
    f_i \equiv f(\params_i),
    \quad 
    w_i \equiv \tilde{\posterior}(\params_i) \Delta \params_i
\end{equation}
then the connection between the weighted sample mean
in equation \eqref{eqn:wt_mean} and the expectation value from our grid
in equation \eqref{eqn:exp_grid} becomes explicit:
\begin{equation}
    \meanwrt{f(\params)}{\posterior}
    \approx \frac{\sum_{i=1}^{n} f(\params_i) 
    \tilde{\posterior}(\params_i) \Delta \params_i}
    {\sum_{i=1}^{n} \tilde{\posterior}(\params_i) \Delta \params_i}
    \equiv \frac{\sum_{i=1}^{n} w_i f_i}
    {\sum_{i=1}^{n} w_i}
\end{equation}

Thinking about our grid as a set of $n$ samples
also allows us to consider an associated \textbf{effective sample size (ESS)}
$n_{\rm eff} \leq n$. The ESS encapsulates the idea that not all of our
samples contribute the same amount of information:
if we have $n$ samples that are very similar to each other, we
expect to have a substantially worse estimate
than if we have $n$ samples that are quite different.
This is because the information in correlated samples are 
at least partially redundant with one another, with the amount of
redundancy increasing with the strength of the correlation: while
two independent samples provide completely unique information about
the distribution and no information about each other, 
two correlated samples instead provide some information about
each other at the expense of the underlying distribution.

\begin{figure}
\begin{center}
\includegraphics[width=\textwidth]{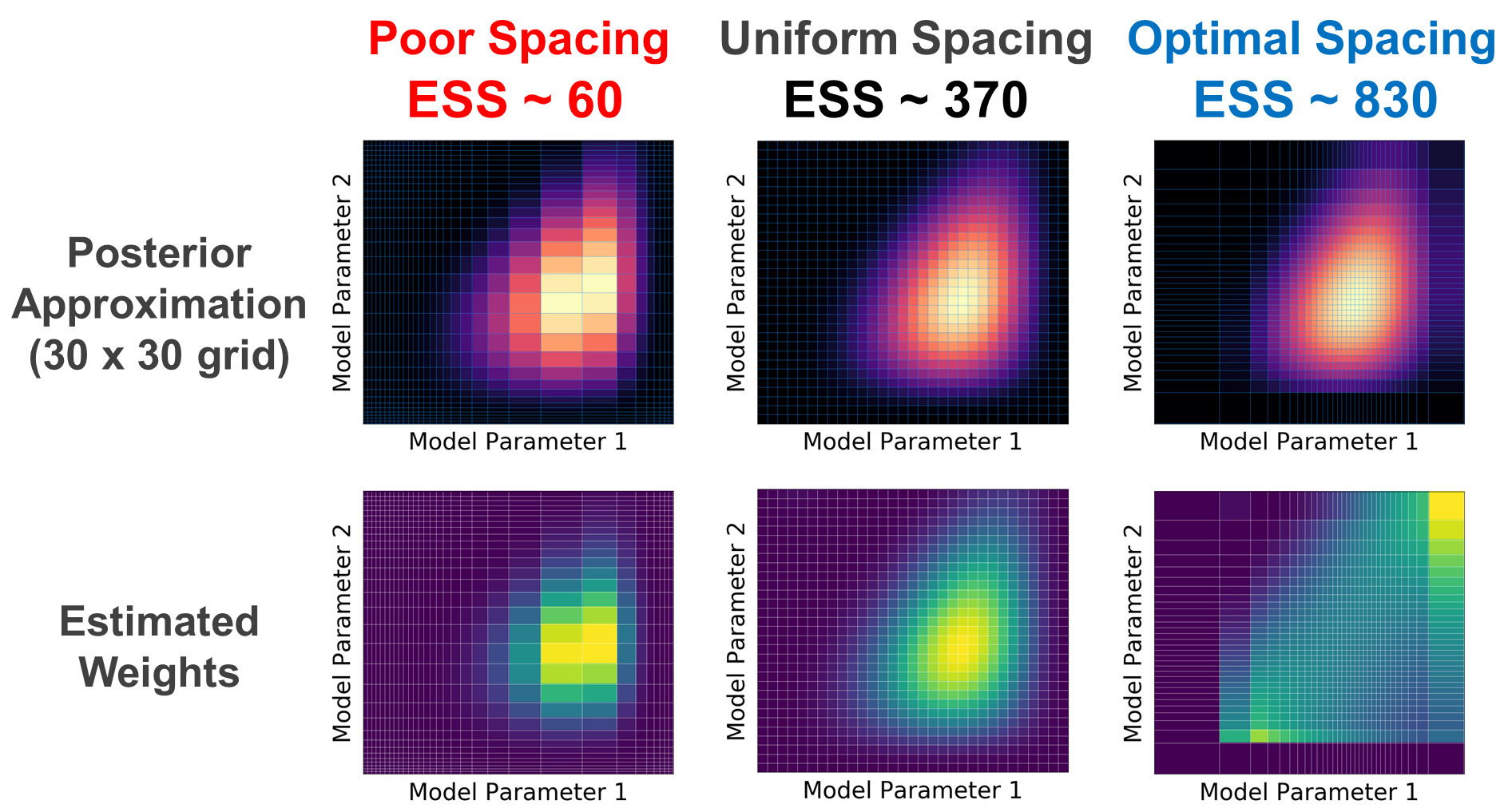}
\end{center}
\caption{An example of how changing the spacing (volume elements) of the grid
can dramatically affect its associated estimate of posterior integrals.
On a toy 2-D posterior $\posterior(\params)$, simply changing the
spacing of the associated 2-D $30 \times 30$ grid dramatically
affects the effective sample size (ESS) (see \S\ref{subsec:ess}).
Differences between poor spacing (left), 
uniform spacing (middle), and optimal spacing (right) leads to
an order of magnitude difference in the ESS, as highlighted by
the distribution of weights (bottom)
associated with the volume elements of each grid.
See \S\ref{subsec:ess} for additional details.
}\label{fig:ess}
\end{figure}

Returning to grids, this correspondence means that we can in theory
come up with an estimate of the expectation value
$\meanwrt{f(\params)}{\posterior}$ that is \textit{at least} as good
as the one we might currently have using a smaller number
$n_{\rm eff} \leq n$ of grid points \textit{if} we were able to allocate them
more efficiently. This distinction matters because errors on 
our estimate of the expectation value generally scale as a function
of $n_{\rm eff}$ rather than $n$. For instance, the error on the
mean typically goes as $\propto n_{\rm eff}^{-1/2}$ rather than
$\propto n^{-1/2}$.

We can quantify the ideas behind the ESS as discussed above
by introducing a formal definition following \citet{kish65}:
\begin{equation}
    n_{\rm eff} 
    \equiv \frac{\left(\sum_{i=1}^{n} w_i\right)^2}{\sum_{i=1}^{n} w_i^2}
\end{equation}
In line with our intuition, the best case under this definition is
one where all the weights are equal ($w_i = w$): 
\begin{equation}
    n_{\rm eff}^{\rm best}
    = \frac{\left(\sum_{i=1}^{n} w_i\right)^2}{\sum_{i=1}^{n} w_i^2}
    = \frac{(nw)^2}{\sum_{i=1}^{n} w^2}
    = \frac{n^2w^2}{nw^2} = n
\end{equation}
Likewise, the worst case is one where
all the weight is concentrated around a single sample 
($w_i = w$ for $i=j$ and $w_i=0$ otherwise):
\begin{equation}
    n_{\rm eff}^{\rm worst}
    = \frac{\left(\sum_{i=1}^{n} w_i\right)^2}{\sum_{i=1}^{n} w_i^2}
    = \frac{(w)^2}{w^2}
    = 1
\end{equation}
This former situation (with $n_{\rm eff}^{\rm best}$)
would be the case where each of the elements of our grid
all have roughly the same contribution to the integral, while the
latter (with $n_{\rm eff}^{\rm worst}$) would be where the entire
integral is essentially contained in just one of our $n$ N-D cuboid regions.
An illustration of this behavior is shown in
{\color{red} \autoref{fig:ess}}.

\subsection{Convergence and Consistency} \label{subsec:consistent}

Now that I have outlined the relationship between the structure
of our grid and the ESS, I want to examine
two final issues: \textbf{convergence} and \textbf{consistency}.
Convergence is the idea that, while our estimates using $n$
samples (grid points) might be noisy, it approaches some fiducial value
as $n \rightarrow \infty$:
\begin{equation}
    \lim_{n \rightarrow \infty} \frac{\sum_{i=1}^{n} f(\params_i) 
    \tilde{\posterior}(\params_i) \Delta \params_i}
    {\sum_{i=1}^{n}\tilde{\posterior}(\params_i) \Delta \params_i}
    = C
\end{equation}
Consistency is subsequently the idea that the value we converge to
is the true value we are interested in estimating:
\begin{equation}
    \lim_{n \rightarrow \infty} \frac{\sum_{i=1}^{n} f(\params_i) 
    \tilde{\posterior}(\params_i) \Delta \params_i}
    {\sum_{i=1}^{n}\tilde{\posterior}(\params_i) \Delta \params_i}
    = \meanwrt{f(\params)}{\posterior}
\end{equation}

\begin{figure}
\begin{center}
\includegraphics[width=\textwidth]{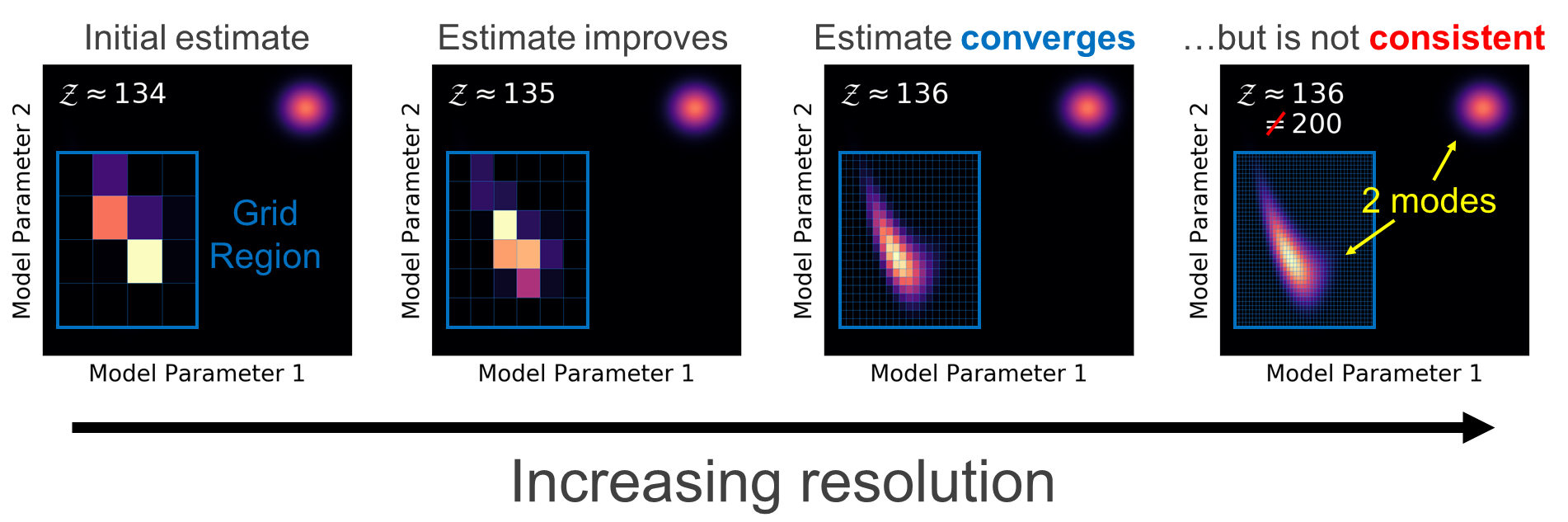}
\end{center}
\caption{An illustration of how grid-based estimates can be \textit{convergent}
(i.e. converge to a single value as the number of grid points increases)
but not \textit{consistent} (i.e. the value it converges to is not the correct answer).
Our toy 2-D unnormalized posterior $\tilde{\posterior}(\params)$ has
two modes that are well-separated with a total evidence of $\evidence = 200$.
If we are not aware of the second mode, we might define a grid region
that only encompasses a subset of the entire parameter space (left). 
While increasing the resolution of the grid within this region allows the estimated
$\evidence$ to converge to an single answer (left to right),
this is not equal to the correct answer of $\evidence = 200$
because we have neglected the contribution of the other component
(right). See \S\ref{subsec:consistent} for additional details.
}\label{fig:conv}
\end{figure}

It is straightforward to show
that \textit{if} the expectation value is well-defined (i.e. it exists)
\textit{and} the grid covers the entire domain of $\params$ (i.e. spans
the smallest and largest possible values in every dimension)
then using a grid is a \textbf{consistent} way to estimate 
the expectation value.
This should make intuitive sense: provided our grid is
expansive enough in $\params$ so that we're not ``missing''
any region of parameter space, we should be able to estimate
$\meanwrt{f(\params)}{\posterior}$ to arbitrary precision
by simply increasing the resolution in $\Delta \params$.

Unfortunately, we do not know beforehand what range of 
values of $\params$ our grid should span. While parameters
can range over $(-\infty, +\infty)$, grids rely on finite-volume elements
and so we have to choose some finite sub-space to grid up.
So while grids may give estimates that converge to some
value over the range spanned by the grid points,
there is always a possibility that a significant portion of the posterior
lies outside that range. In these cases,
grids are not guaranteed to be consistent estimators
of $\meanwrt{f(\params)}{\posterior}$. An illustration of this issue is
shown in {\color{red} \autoref{fig:conv}}.
This fundamental problem is not shared by Monte Carlo
methods, which I will cover in \S\ref{sec:montecarlo}.

\subsection*{Exercise: Grids over a 2-D Gaussian} \label{exercise:grids}

\subsubsection*{Setup}

Consider an unnormalized posterior 
well-approximated by a 2-D Gaussian (Normal)
distribution centered on $(\mu_x,\mu_y)$ 
with standard deviations $(\sigma_x, \sigma_y)$:
\begin{equation*}
    \tilde{\posterior}(x,y) 
    = \exp\left\{-\frac{1}{2}\left[\frac{(x-\mu_x)^2}{\sigma_x^2}
    + \frac{(y-\mu_y)^2}{\sigma_y^2}\right]\right\}
\end{equation*}
Assume that we expect to find our posterior has a mean of $0$ and
a standard deviation of $1$.
In reality, however, our posterior actually has means $(\mu_x,\mu_y)=(-0.3,0.8)$ and
standard deviations $(\sigma_x^2,\sigma_y^2)=(2,0.5)$, mimicking the common
case where our prior expectations and posterior inferences somewhat disagree.

\subsubsection*{Grid-based Estimation}

We want to use a 2-D grid to estimate various forms of posterior integrals.
Starting with an evenly-spaced $5 \times 5$ grid from $[-2, 2]$, compute:
\begin{enumerate}
    \item the evidence $\evidence$,
    \item the means $\meanwrt{x}{\posterior}$
    and $\meanwrt{y}{\posterior}$,
    \item the 68\% credible intervals (or closest approximation) 
    $[x_{\rm low}, x_{\rm high}]$ and $[y_{\rm low}, y_{\rm high}]$,
    \item and the effective sample size $n_{\rm eff}$.
\end{enumerate}
How accurate are each of these quantities with the values we might
expect? What does $n_{\rm eff}/n$ tell us about how efficiently
we have allocated our grid points?

\subsubsection*{Convergence}

Repeat the above exercise using an evenly-spaced grid of
$20 \times 20$ points and $100 \times 100$ points. Comment
on any differences. How much has the overall accuracy improved? Do
the estimates appear convergent?

\subsubsection*{Consistency}

Next, expand the bounds of the grid to be from $[-5, 5]$ and
perform the same exercise as above. Do the answers change substantially?
If so, what does this tell us about the consistency of our previous
estimates? Adjust the density and bounds of the grid until the 
answers appear both convergent and consistent.
Remember that we do not know the exact shape of the posterior 
ahead of time. What does this imply
about general concerns when applying grids in practice?

\subsubsection*{Effective Sample Size}

Finally, explore whether there is a straightforward scheme to adjust the
locations of the $x$ and $y$ grid points to maximize the effective
sample size based on the definition outlined in \S\ref{subsec:ess}.
If so, can you explain why it works? If not, why not?
Compared to equivalent evenly-spaced grids, how much can adaptively adjusting
the grid spacing improve $n_{\rm eff}$ and the overall accuracy of our estimates?

\section{From Grids to Monte Carlo Methods} \label{sec:montecarlo}

\subsection{Connecting Grid Points and Samples}
\label{subsec:grid_to_samp}

Earlier, I outlined how we can relate estimating
$\meanwrt{f(\params)}{\posterior}$ using a grid of $n$ points to
an equivalent estimate using a set of $n$ samples $\{ f_1, \dots, f_n\}$
and a series of associated weights $\{ w_1, \dots, w_n \}$.
The main result is that there is an intimate connection between the
structure of the posterior and the grid to the relative amplitude
of the weights $w_i \equiv \tilde{\posterior}(\params_i)\Delta\params_i$
for each point $f_i \equiv f(\params_i)$. Adjusting the resolution
of the grid then affects these weights, with a more uniform distribution
of weights leading to a larger ESS which can improve our estimate.

The fact that decreasing the spacing (making grid denser)
also decreases the weights makes sense: we have more points 
located in that region, so each point should in general
get less relative weight when computing $\meanwrt{f(\params)}{\posterior}$. 
Likewise, if we have the same spacing
but change the relative shape of the posterior, 
the weight of that point when estimating $\meanwrt{f(\params)}{\posterior}$
should also change accordingly.

I now want to extend this basic relationship further.
In theory, adaptively increasing the resolution of our grid allows
us more control over the volume elements $\Delta \params_i$ used to
derive our weights. If we knew the shape of our posterior sufficiently
well, for large $n$ 
we should in theory be able to adjust $\Delta \params_i$ such that
the weights $w_i = \tilde{\posterior}(\params_i)\Delta\params_i$
are uniform to some amount of desired precision.
By inspection, this should happen when
\begin{equation}
    \Delta \params_i \propto \frac{1}{\tilde{\posterior}(\params_i)}
\end{equation}
for all $i$.

\begin{figure}
\begin{center}
\includegraphics[width=\textwidth]{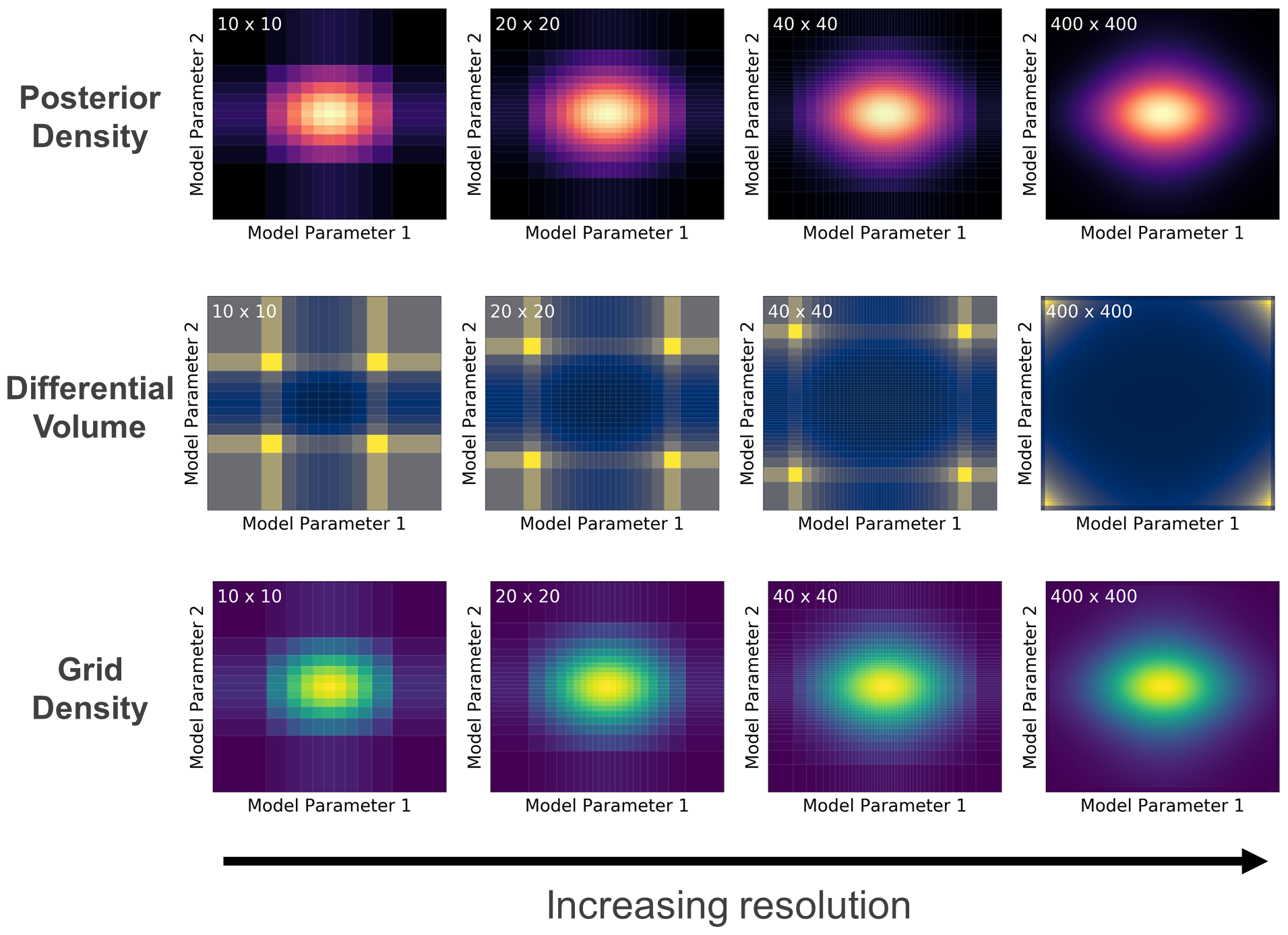}
\end{center}
\caption{An illustration of the connection between grids and
continuous density distributions. As we increase the number
of grid points, our estimate of the posterior $\posterior(\params)$
improves (top). Since the spacing between the grid points varies
to maximize the effective sample size (see \autoref{fig:ess} and
\S\ref{subsec:ess}), the differential volume elements $\Delta \params_i$
change depending on our location (middle). As we continue to
increase the number of volume elements, the density of grid
points at any particular location 
$\rho(\params_i) = [\Delta \params_i]^{-1}$ 
behaves like a continuous function $\proposal(\params)$
whose distribution is similar to $\posterior(\params)$ (bottom).
This implies we should be able to use $\proposal(\params)$
in some way to estimate $\posterior(\params)$.
See \S\ref{sec:montecarlo} for additional details.
}\label{fig:density}
\end{figure}

Taking this reasoning to its conceptual limit, as $n \rightarrow \infty$
we can imagine estimating the posterior using a larger and larger number of 
grid points whose spacing $\Delta \params$ changes as a function of $\params$.
Using this, we can now define the density of points $\proposal(\params)$
based on the varying resolution $\Delta\params(\params)$ of our
infinitely-fine grid as a function of $\params$:
\begin{equation}
    \proposal(\params) \propto \frac{1}{\Delta\params (\params)}
\end{equation}
This result suggests that, in the continuum limit where $n \rightarrow \infty$, 
\textit{the structure of our infinite-resolution grid is equivalent to a new
continuous distribution} $\proposal(\params)$. An illustration
of this concept is shown in {\color{red} \autoref{fig:density}}. 
Using $\proposal(\params)$, we can 
then rewrite our original expectation value as
\begin{equation}
    \meanwrt{f(\params)}{\posterior} 
    \equiv \frac{\int f(\params) \tilde{\posterior}(\params) \deriv \params}
    {\int \tilde{\posterior}(\params) \deriv \params}
    = \frac{\int f(\params) \frac{\tilde{\posterior}(\params)}{\proposal(\params)}
    \proposal(\params) \deriv \params}
    {\int \frac{\tilde{\posterior}(\params)}{\proposal(\params)}
    \proposal(\params) \deriv \params}
    = \frac{\meanwrt{f(\params) 
    \tilde{\posterior}(\params)/\proposal(\params)}{\proposal}}
    {\meanwrt{\tilde{\posterior}(\params)/\proposal(\params)}{\proposal}}
\end{equation}
For reasons that will soon become clear, I will refer to 
$\proposal(\params)$ as the \textbf{proposal distribution}.

At this point, this may mostly seem like a mathematical trick:
all I have done is rewrite our original \textit{single} 
expectation value with respect to the (unnormalized)
posterior $\tilde{\posterior}(\params)$ in terms of \textit{two}
expectation values with respect to the proposal distribution
$\proposal(\params)$. 
This substitution, however, actually allows us to fully realize 
the connection between grid points and samples.

Earlier, I showed that the estimate for the expectation
value from grid points is exactly analogous to the estimate
we would derive assuming the grid points were random samples
$\{ f_1, \dots, f_n \}$ with associated weights
$\{ w_1, \dots, w_n \}$. Once we have defined our
expectation with respect to $\proposal(\params)$, however,
this statement can become exact assuming we can explicitly generate
samples from $\proposal(\params)$.

Let's quickly review what this means. Initially,
we looked at trying to estimate $\meanwrt{f(\params)}{\posterior}$ over a grid
with $n$ points. In the limit of infinite resolution, however, our grid becomes
equivalent to some distribution $\proposal(\params)$. Using $\proposal(\params)$,
we can then rewrite our original expression in terms of two expectations,
$\meanwrt{f(\params) \tilde{\posterior}(\params)/\proposal(\params)}{\proposal}$
and $\meanwrt{\tilde{\posterior}(\params)/\proposal(\params)}{\proposal}$, over
$\proposal(\params)$ instead of $\posterior(\params)$.
This helps us because we can in theory estimate these final expressions
explicitly using a series of $n$ randomly generated samples from
$\proposal(\params)$. Due to the randomness inherent in this approach, 
this is commonly referred to as a \textbf{Monte Carlo} approach for estimating
$\meanwrt{f(\params)}{\posterior}$ due to historical connections with
randomness and gambling.

On the face of it, this should come across as a surprising claim. When we
compute an integral of a function $f(\params)$ on a bounded grid, we
know that there is some error in our approximation having to do with the
discretization of the grid. This error is entirely \textit{deterministic}:
given a number of grid points $n$ and an a particular discretization density
$\proposal(\params) \propto 1/\Delta\params(\params)$, we will get the
same result (and error) for $\meanwrt{f(\params)}{\posterior}$ every time.

By contrast, drawing $n$ samples $\{\params_1,\dots,\params_n\}$
from $\proposal(\params)$ is an inherently \textit{random} (i.e. stochastic) 
process that seems to look nothing like a
grid of points. And because these points are inherently random,
the actual deviation between our estimate and the true value of
$\meanwrt{f(\params)}{\posterior}$ will \textit{also} be random.
The ``error'' from random samples then tells us something about
how much we expect our estimate can differ over many possible
realizations of our random process given a particular number of samples
$n$ generated from $\proposal(\params)$. The fact that we can derive
roughly equivalent estimates from these these very different
approaches as we adjust $n$ and $\proposal(\params)$ lies at the heart of the
connection between grid points and samples.

There are three primary benefits from moving from an adaptively-spaced grid
to a continuous distribution $\proposal(\params)$.
First, a grid will always have some minimum 
resolution $\Delta \params_i$ that makes it difficult
to get our weights to be roughly uniform, limiting our maximum
ESS in practice. By contrast, we can in theory get
$\proposal(\params)$ to more closely match the posterior $\posterior(\params)$,
giving a larger ESS at fixed $n$.

Second, because we are now working with \textit{distributions}
rather than a finite number of grid points, we are no longer limited
to some finite volume when estimating expectations. Since distributions can
range over $(-\infty, +\infty)$, we can guarantee $\proposal(\params)$
will provide sufficient \textbf{coverage} over all possible $\params$ 
values that our posterior $\posterior(\params)$ could be defined over.
This means that some of the theoretical issues raised in \S\ref{subsec:consistent}
associated with applying grids to posteriors 
that range over $(-\infty, +\infty)$ no longer apply. 
Monte Carlo methods therefore can serve as a \textit{consistent}
estimator for a wider range of possible posterior expectations
than grid-based methods, making them substantially more flexible.

Finally, the minimum number of grid points always scales exponentially
with dimensionality (see \S\ref{subsec:curse}), regardless of how many 
parameters we are interested in marginalizing over. Since 
Monte Carlo methods do not rely on these, they can take full
advantage of marginalizing over parameters 
when estimating expectations $\meanwrt{f(\params)}{\posterior}$. 
They are therefore less susceptible to this effect 
(although see \S\ref{subsec:volume}).

\subsection{Importance Sampling} \label{subsec:importance}

As I have tried to emphasize previously, the core tenet of this article is that
\textit{we do not know what $\posterior(\params)$ looks like beforehand}.
This means we do not know what grid structure will
provide an optimal estimate (i.e. maximum ESS) for
$\meanwrt{f(\params}{\posterior}$, let alone how this should
behave as $\proposal(\params)$ in the continuum limit. This gives
us ample motivation to \textit{choose} $\proposal(\params)$
in such a way to make generating samples from it easy and straightforward.

Assuming we have chosen such a $\proposal(\params)$, we can 
subsequently generate a series of $n$ samples from it. Assuming
these samples have weights $q_i$ associated with them and defining
\begin{equation}
    f(\params_i) \equiv f_i, \quad
    \tilde{\posterior}(\params_i)/\proposal(\params_i) 
    \equiv \tilde{w}(\params_i) \equiv \tilde{w}_i
\end{equation}
our original expression reduces to
\begin{equation}
    \meanwrt{f(\params)}{\posterior} 
    = \frac{\meanwrt{f(\params) 
    \tilde{w}(\params)}{\proposal}}
    {\meanwrt{\tilde{w}(\params)}{\proposal}}
    \approx \frac{\sum_{i=1}^{n} f_i \tilde{w}_i q_i}
    {\sum_{i=1}^{n} \tilde{w}_i q_i}
\end{equation}
If we further assume that we have chosen $\proposal(\params)$ so that
we can simulate samples that are 
\textbf{independently and identically distributed (iid)}
(i.e. each sample has the same probability distribution 
as the others and all the samples are mutually independent),
then the corresponding sample weights immediately reduce to 
$q_i = 1/n$ and our result becomes
\begin{equation}
    \meanwrt{f(\params)}{\posterior} 
    \approx \frac{n^{-1} \sum_{i=1}^{n} f_i \tilde{w}_i}
    {n^{-1} \sum_{i=1}^{n} \tilde{w}_i}
\end{equation}
As with the previous case using grids (\S\ref{sec:grid}),
the denominator of this
expression is again a direct approximation for the evidence
\begin{align}
    \evidence 
    = \int \tilde{\posterior}(\params) \deriv \params
    \approx n^{-1} \sum_{i=1}^{n} \tilde{w}_i
\end{align}

\begin{figure}
\begin{center}
\includegraphics[width=\textwidth]{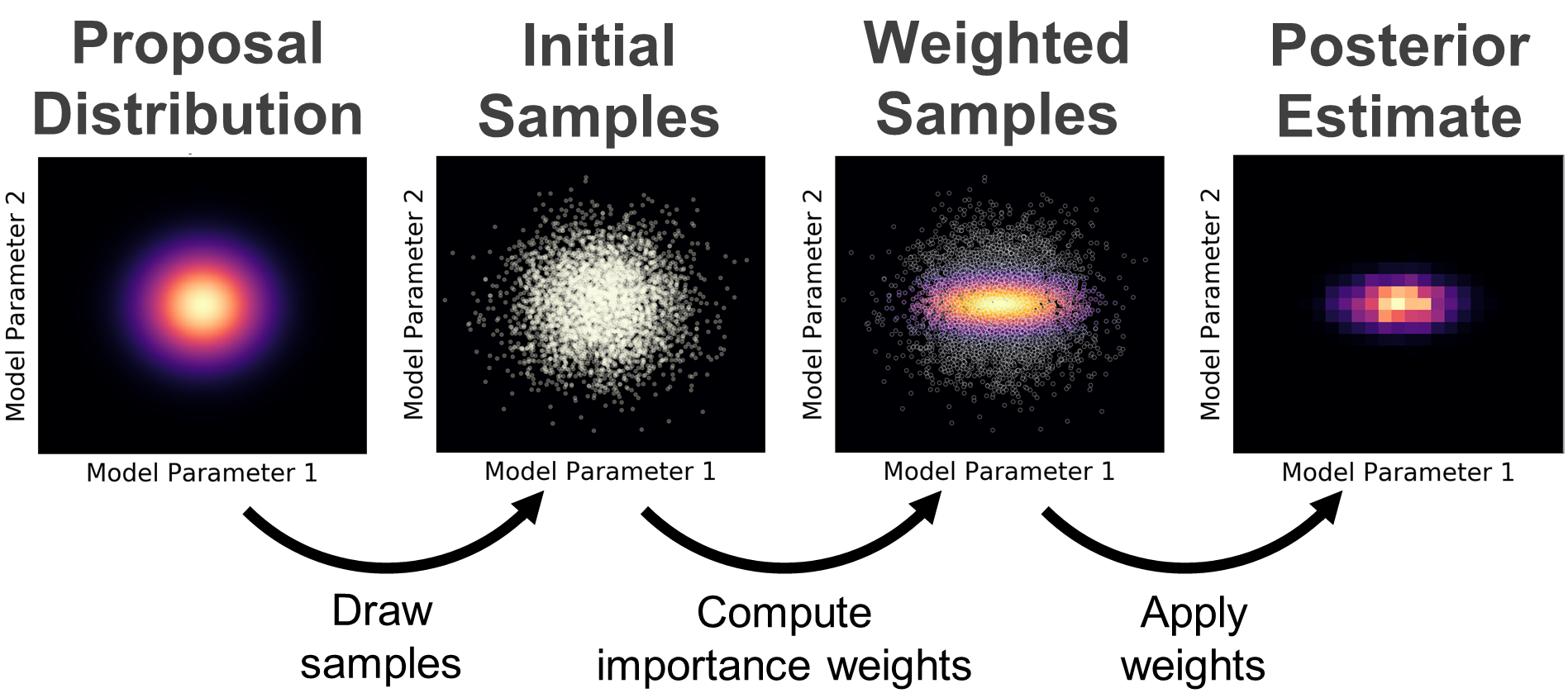}
\end{center}
\caption{A schematic illustration of Importance Sampling.
First, we take a given proposal distribution 
$\proposal(\params)$ (left) and generate a set of $n$ iid samples
from it (middle left).
We then weight each sample based on the corresponding
``importance'' $\tilde{\posterior}(\params)/\proposal(\params)$
it has at that location (middle right). We then can use these
weighted samples to approximate posterior expectations
(right).
See \S\ref{subsec:importance} for additional details.
}\label{fig:importance}
\end{figure}

This gives a straightforward recipe for estimating our original expectation
value:
\begin{enumerate}
    \item Draw $n$ iid samples $\{\params_1, \dots, \params_n \}$
    from $\proposal(\params)$.
    \item Compute their corresponding weights $\tilde{w}_i =
    \tilde{\posterior}(\params_i)/\proposal(\params_i)$.
    \item Estimate $\meanwrt{f(\params)}{\posterior}$
    by computing $\meanwrt{\tilde{w}(\params)}{\proposal}$ and
    $\meanwrt{f(\params)\tilde{w}(\params)}{\proposal}$ using the
    weighted sample means.
\end{enumerate}
Since this process just involves ``reweighting'' the samples
based on $\tilde{w}_i$, these weights are often referred to as
\textbf{importance weights} and the method as \textbf{Importance Sampling}.
A schematic illustration of Importance Sampling
is highlighted in {\color{red} \autoref{fig:importance}}.

We can interpret the importance weights as ways to correct for how
``far off'' our original guess $\proposal(\params)$ 
is from the truth $\posterior(\params)$.
If the posterior density is higher at position
$\params_i$ relative to the proposal density, 
then we were less likely to generate
a sample at that position compared to what we would have
seen if we had drawn samples directly from the posterior.
As a result, we should increase its corresponding weight
to account for this expected deficit of samples at a given position.
If the posterior density is lower relative to the proposal density, 
then the alternative is true and we want to lower the weight of the
corresponding sample to account for the expected excess of samples
at a given position.

\subsection{Examples of Sampling Strategies} \label{subsection:samp_strat}

Importance Sampling serves as a useful first step for understanding how
the weights $\{ \tilde{w}_1, \dots, \tilde{w}_n \}$ for the corresponding set 
of $n$ samples are related to different Monte Carlo sampling strategies.

As an example, one common approach is to generate samples uniformly within some 
cuboid with volume $V$. The proposal distribution for this
will then be
\begin{equation}
    \proposal^{\rm unif}(\params) =
    \begin{cases}
    1/V & \params \: {\rm in\:cuboid} \\
    0 & {\rm otherwise}
    \end{cases}
\end{equation}
The corresponding importance weights subsequently will
just be proportional to the posterior at a given position:
\begin{equation}
    \tilde{w}_i^{\rm unif}
    = \frac{\tilde{\posterior}(\params_i)}{\proposal^{\rm unif}(\params_i)}
    = V \tilde{\posterior}(\params_i)
    \propto \posterior(\params_i)
\end{equation}

Another possible approach would be if we instead take our proposal
to be our prior:
\begin{equation}
    \proposal^{\rm prior}(\params) = \prior(\params)
\end{equation}
This seems like a well-motivated choice: the prior characterizes
our knowledge before looking at the data, so it should serve
as a useful first guess and encompass the range of all possibilities.
Under this assumption, we now find our weights will be 
equal to the likelihood $\likelihood(\params)$ at each position:
\begin{equation}
    w_i^{\rm prior} 
    = \frac{\tilde{\posterior}(\params_i)}{\proposal^{\rm prior}(\params_i)}
    = \frac{\likelihood(\params_i) \prior(\params_i)}{\prior(\params_i)}
    = \likelihood(\params_i)
\end{equation}

Finally, notice that the optimal sampling strategy is to
assume that we can take our proposal to be identical to our
posterior:
\begin{equation}
    \proposal^{\rm post}(\params) = \posterior(\params)
\end{equation}
The corresponding weights will then just be constant and
equal to the evidence $\evidence$:
\begin{equation}
    w_i^{\rm post} 
    = \frac{\tilde{\posterior}(\params_i)}{\proposal^{\rm post}(\params_i)}
    = \frac{\evidence \posterior(\params_i)}{\posterior(\params_i)}
    = \evidence
\end{equation}

As expected, this final result guarantees the maximum
possible ESS of $n_{\rm eff} = n$. Getting
$\proposal(\params)$ to be as ``close'' as possible
to $\posterior(\params)$ therefore becomes a crucial part of
analyses when trying to use Importance Sampling 
to estimate expectation values. It is this result in particular 
that motivates the use of Markov Chain Monte Carlo (MCMC) methods
discussed from \S\ref{sec:mcmc} onward:
if we can somehow generate samples \textit{directly}
from $\posterior(\params)$ or something close to it, 
then we can achieve an optimal estimate of our
corresponding expectation values.

\subsection*{Exercise: Importance Sampling over a 2-D Gaussian} 
\label{exercise:importance}

\subsubsection*{Setup}

Let's return to our exercise from \S\ref{sec:grid}, in which
our unnormalized posterior is well-approximated by a 2-D Gaussian (Normal)
distribution:
\begin{equation*}
    \tilde{\posterior}(x,y) 
    = \exp\left\{-\frac{1}{2}\left[\frac{(x-\mu_x)^2}{\sigma_x^2}
    + \frac{(y-\mu_y)^2}{\sigma_y^2}\right]\right\}
\end{equation*}
where $(\mu_x,\mu_y)=(-0.3,0.8)$ and $(\sigma_x^2,\sigma_y^2)=(2,0.5)$.

\subsubsection*{Importance Sampling}

We want to use Importance Sampling to approximate various posterior
integrals from this distribution.
We will start by choosing our proposal distribution $\proposal(x,y)$
to be a 2-D Gaussian with a mean of $0$ and standard deviation of $1$:
\begin{equation*}
    \proposal(x,y) = \Normal{(\mu_x,\mu_y)=(0,0)}{(\sigma_x,\sigma_y)=(1,1)}
\end{equation*}

Using $n=25$ iid random samples drawn 
from the proposal distribution, compute an estimate for:
\begin{enumerate}
    \item the evidence $\evidence$,
    \item the means $\meanwrt{x}{\posterior}$
    and $\meanwrt{y}{\posterior}$,
    \item the 68\% credible intervals (or closest approximation) 
    $[x_{\rm low}, x_{\rm high}]$ and $[y_{\rm low}, y_{\rm high}]$,
    \item and the effective sample size $n_{\rm eff}$.
\end{enumerate}
How accurate are each of these quantities with the values we might
expect? What does $n_{\rm eff}/n$ tell us about how well
our proposal $\proposal(x,y)$ traces the underlying posterior
$\posterior(x,y)$?

\subsubsection*{Uncertainty}

Repeat the above exercise $m=100$ times to get an
estimate for how much our estimates of each quantity can vary.
Is the variation in line with what might be expected given 
the typical effective sample size? Why or why not?

\subsubsection*{Convergence}

Now repeat the above exercise using $n=100$, $n=1000$, and $n=10000$ 
points rather than $n=25$ points and comment on any differences.
How much has the overall accuracy improved? Do
the estimates appear convergent and consistent as $n_{\rm eff}$ increases?
How much do the errors on quantities shrink as a function
of $n$ and/or $n_{\rm eff}$? Is this behavior expected? Why or why not?

\subsubsection*{Consistency}

Next, let's expand our proposal distribution to instead have
$(\sigma_x,\sigma_y)=(2,2)$ to get more coverage in the ``tails'' of the
posterior. Perform the same exercise as above 
with $n=\{100,1000,10000\}$ iid random samples. 
Do the answers change substantially? Why or why not?

While in theory we can choose $\proposal(x,y) \approx \posterior(x,y)$ 
so that $n_{\rm eff} \approx n$, we do not know the exact shape of the posterior 
ahead of time. Given that $\tilde{\posterior}(x,y)$ may differ from
our initial expectations, what does this exercise imply about general concerns
applying Importance Sampling in practice?

\section{Markov Chain Monte Carlo} \label{sec:mcmc}

Now that we see how the weights relate to various Monte Carlo sampling strategies
(e.g., generating samples from the prior), I will now outline
the idea behind \textbf{Markov Chain Monte Carlo (MCMC)}. In brief,
MCMC methods try to generate samples in such a way that the importance
weights $\{ \tilde{w}_1, \dots, \tilde{w}_n \}$ associated with each sample
are constant. Based on the results from \S\ref{subsection:samp_strat}, this
means MCMC seeks to generate samples proportional to
the posterior $\posterior(\params)$ in order to arrive 
at an \textit{optimal estimate} for our expectation value.

\begin{figure}
\begin{center}
\includegraphics[width=\textwidth]{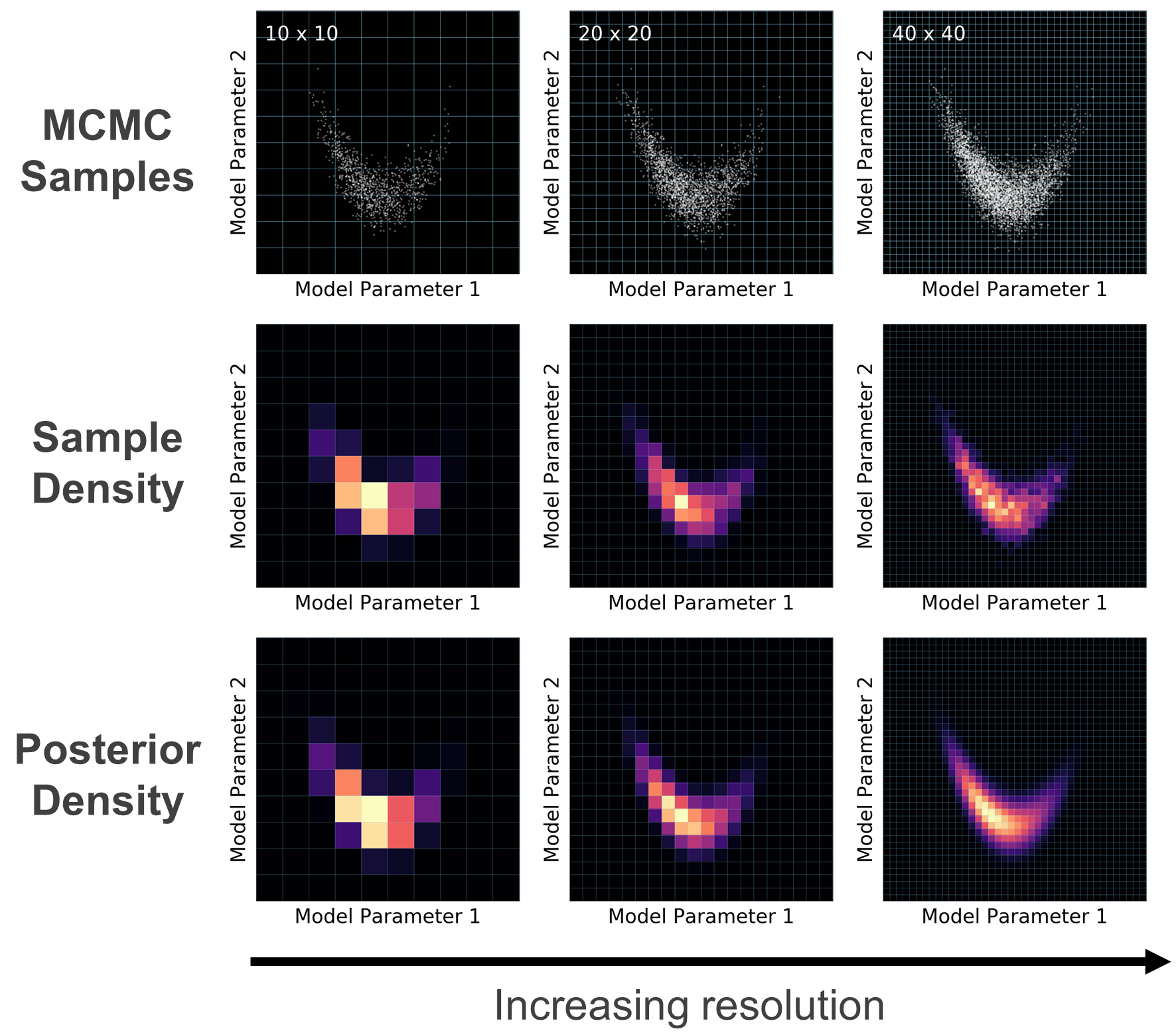}
\end{center}
\caption{A schematic illustration of Markov Chain Monte Carlo (MCMC).
MCMC tries to create a chain of $n$ (correlated) samples 
$\{ \params_1 \rightarrow \dots \rightarrow \params_n \}$ (top)
such that the number of samples $m$ in some particular volume
$\delta$ gives a relative density $m/n$ (middle) comparable to the
posterior $\posterior(\params)$ integrated over the same volume (bottom).
See \S\ref{sec:mcmc} for additional details.
}\label{fig:mcmc}
\end{figure}

MCMC accomplishes this by creating a \textbf{chain} of 
(correlated) parameter values 
$\{ \params_1 \rightarrow \dots \rightarrow \params_n \}$
over $n$ iterations such that the number of iterations $m(\params_i)$
spent in any particular region $\delta_{\params_i}$ 
centered on $\params_i$ is proportional to the posterior 
density $\posterior(\params_i)$ contained within that region.
In other words, the ``density'' of samples generated from MCMC
\begin{equation}
    \rho(\params) \equiv \frac{m(\params)}{n}
\end{equation}
at position $\params$ integrated over $\delta_{\params}$ is approximately
\begin{equation}
    \int_{\params \in \delta_{\params}} \posterior(\params) \deriv \params
    \approx \int_{\params \in \delta_{\params}} \rho(\params) \deriv \params 
    \approx n^{-1} \sum_{j=1}^{n} \indicator{\params_j \in \delta_{\params}}
\end{equation}
where $\indicator{\cdot}$ is the \textbf{indicator function} which evaluates to
$1$ if the inside condition is true and $0$ otherwise. We can therefore
approximate the density by simply
adding up the number of samples within $\delta_{\params}$
and normalizing by the total number of samples $n$.
A schematic illustration of this concept is shown in
{\color{red} \autoref{fig:mcmc}}.

While this will just be approximately true for any finite $n$,
as the number of samples $n \rightarrow \infty$ this procedure generally
guarantees that $\rho(\params) \rightarrow \posterior(\params)$ 
everywhere.\footnote{Discussing the details of exactly when/where this
condition holds in theory and in practice
is beyond the scope of this paper but can be found in
other references such as \citet{asmussenglynn11} and \citet{brooks+11}.}
In theory then, once we have a reasonable enough
approximation for $\rho(\params)$, we can also use the 
samples $\{ \params_1 \rightarrow \dots \rightarrow \params_n \}$
generated from $\rho(\params)$
to get an estimate for the evidence using the 
same substitution trick introduced in \S\ref{sec:montecarlo}:
\begin{align}
    \evidence
    = \int \frac{\tilde{\posterior}(\params)}{\rho(\params)}
    \rho(\params) \deriv \params
    \equiv \meanwrt{\tilde{\posterior}(\params)/\rho(\params)}{\rho}
    \approx n^{-1} \sum_{i=1}^{n} 
    \frac{\tilde{\posterior}(\params_i)}{\rho(\params_i)}
\end{align}
This is just the average of the ratio 
between $\tilde{\posterior}(\params_i)$ and $\rho(\params_i)$
over all $n$ samples.

Finally, since our MCMC procedure gives us a series of $n$
samples from the posterior, our expectation value simply reduces to
\begin{equation}
    \meanwrt{f(\params)}{\posterior} 
    \approx \frac{n^{-1} \sum_{i=1}^{n} f_i \tilde{w}_i}
    {n^{-1} \sum_{i=1}^{n} \tilde{w}_i}
    = \frac{n^{-1} \sum_{i=1}^{n} f_i}
    {n^{-1} \sum_{i=1}^{n} 1}
    = n^{-1} \sum_{i=1}^{n} f_i
\end{equation}
This is just the \textbf{sample mean} of
the corresponding $\{ f_1, \dots, f_n \}$ values over
our set of $n$ samples. 

I wish to take a moment here to highlight two features of the above results
related to common misconceptions surrounding MCMC methods. First, there
is a widespread belief that because MCMC methods generate a chain of 
samples whose behavior \textit{follows} the posterior, we do not
have any ability to use them to estimate normalizing constants such
as the evidence $\evidence$. As shown above, this is not true at all: not only
\textit{can} we do this using $\rho(\params)$, but the estimate we derive
is actually a \textit{consistent} one (although it will converge slowly;
see \S\ref{subsec:post_approx}).

The second misconception is that the primary goal of MCMC is to
``approximate'' or ``explore'' the posterior. In other words,
to estimate $\rho(\params)$. However, as shown above,
the ability of MCMC methods to estimate $\rho(\params)$
is really only useful for estimating the evidence $\evidence$. In fact, by tracing
its heritage from Importance Sampling-based methods, we see its primary purpose
is actually \textit{to estimate expectation values} (i.e.
integrals \textit{over} the posterior). I have explicitly tried to avoid
introducing any mention of ``approximating the posterior'' up to this point
in order to avoid this misconception, but will spend some time discussing
this point in more detail in \S\ref{subsec:post_approx}.

To summarize, the idea behind MCMC is to simulate a series of
values $\{ \params_1 \rightarrow \dots \rightarrow \params_n \}$
in a way that their density $\rho(\params)$ after a given amount
of time follows the underlying posterior $\posterior(\params)$. We can
then estimate the posterior within any particular region $\delta_{\params}$
by simply counting up how many samples we simulate there and normalizing by
the total number of samples $n$ we generated. Because we are also
simulating values directly from the posterior, any expectation values
also reduce to simple sample averages. This procedure is incredibly
intuitive and part of the reason MCMC methods have 
become so widely adopted.

\subsection{Generating Samples with the Metropolis-Hastings Algorithm} \label{subsec:mh}

There is a vast literature on various approaches to generating samples
\textbf{(see, e.g., cites)}. Since this article focuses on
building up a \textit{conceptual understanding} of MCMC methods, 
exploring how the majority of these methods behave both in theory and
in practice is beyond the scope of this paper.

Instead of an overview, I aim to clarify the basics of how these methods operate.
The central idea is that we want a way to generate new samples 
$\params_i \rightarrow \params_{i+1}$ such that the
distribution of the final samples $\rho(\params)$ as $n \rightarrow \infty$
(1) is \textbf{stationary} (i.e. it converges to something) and 
(2) is equal to the $\posterior(\params)$. These are essentially analogs
to the convergence and consistency constraints 
discussed in \S\ref{subsec:consistent}.

We can satisfy the first condition by invoking \textbf{detailed balance}.
This is the idea that probability is conserved when moving
from one position to another (i.e. the process is reversible). More formally,
this just reduces to factoring of probability:
\begin{equation}
    P(\params_{i+1}|\params_i) P(\params_i) 
    = P(\params_{i+1}, \params_i)
    = P(\params_i|\params_{i+1}) P(\params_{i+1}) 
\end{equation}
where $P(\params_{i+1}|\params_i)$ is the probability of moving
from $\params_i \rightarrow \params_{i+1}$ and $P(\params_{i}|\params_{i+1})$ 
is the probability of the reverse move from 
$\params_{i+1} \rightarrow \params_i$. 
Rearranging then gives the following constraint:
\begin{equation}
    \frac{P(\params_{i+1}|\params_i)}{P(\params_i|\params_{i+1})} 
    = \frac{P(\params_{i+1})}{P(\params_i)}
    = \frac{\posterior(\params_{i+1})}{\posterior(\params_i)}
\end{equation}
where the final equality comes from the fact that the distribution
we are trying to generate samples from is the posterior $\posterior(\params)$.

We now need to implement a procedure that enables us to actually
move to new positions by computing this probability. 
We can do this by breaking each move into two steps. First,
we want to \textit{propose} a new position 
$\params_i \rightarrow \params_{i+1}'$ based on a
\textbf{proposal distribution} $\proposal(\params_{i+1}'|\params_i)$
similar in nature to the $\proposal(\params)$ used in to 
Importance Sampling (\S\ref{subsec:importance}). 
Then we will either decide to \textbf{accept} the new position
($\params_{i+1}=\params_{i+1}'$) or \textbf{reject} the new position
($\params_{i+1}=\params_i$) with some \textbf{transition probability}
$T(\params_{i+1}'|\params_i)$.
Combining these terms together then gives us the probability
of moving to a new position:
\begin{equation}
    P(\params_{i+1}|\params_i) 
    \equiv \proposal(\params_{i+1}|\params_i) T(\params_{i+1}|\params{i})
\end{equation}

As with Importance Sampling, we can choose $\proposal(\params_{i+1}'|\params_i)$
so that it is straightforward to propose new samples
$\params_{i+1}'$ by numerical simulation. 
We then need to determine the transition probability
$T(\params_{i+1}'|\params_i)$ of whether we should accept 
or reject $\params_{i+1}'$. Substituting into
our expression for detailed balance, we find that our form for the
transition probability must satisfy the following constraint:
\begin{equation}
    \frac{T(\params_{i+1}|\params_i)}{T(\params_i|\params_{i+1})} 
    = \frac{\posterior(\params_{i+1})}{\posterior(\params_i)}
    \frac{\proposal(\params_i|\params_{i+1})}{\proposal(\params_{i+1}|\params_i)}
\end{equation}
It is straightforward to show that the 
\textbf{Metropolis criterion} \cite{metropolis+53_alt}
\begin{equation}
    T(\params_{i+1}|\params_i)
    \equiv \min\left[1, \frac{\posterior(\params_{i+1})}{\posterior(\params_i)}
    \frac{\proposal(\params_i|\params_{i+1})}{\proposal(\params_{i+1}|\params_i)}
    \right]
\end{equation}
satisfies this constraint.

\begin{figure}
\begin{center}
\includegraphics[width=\textwidth]{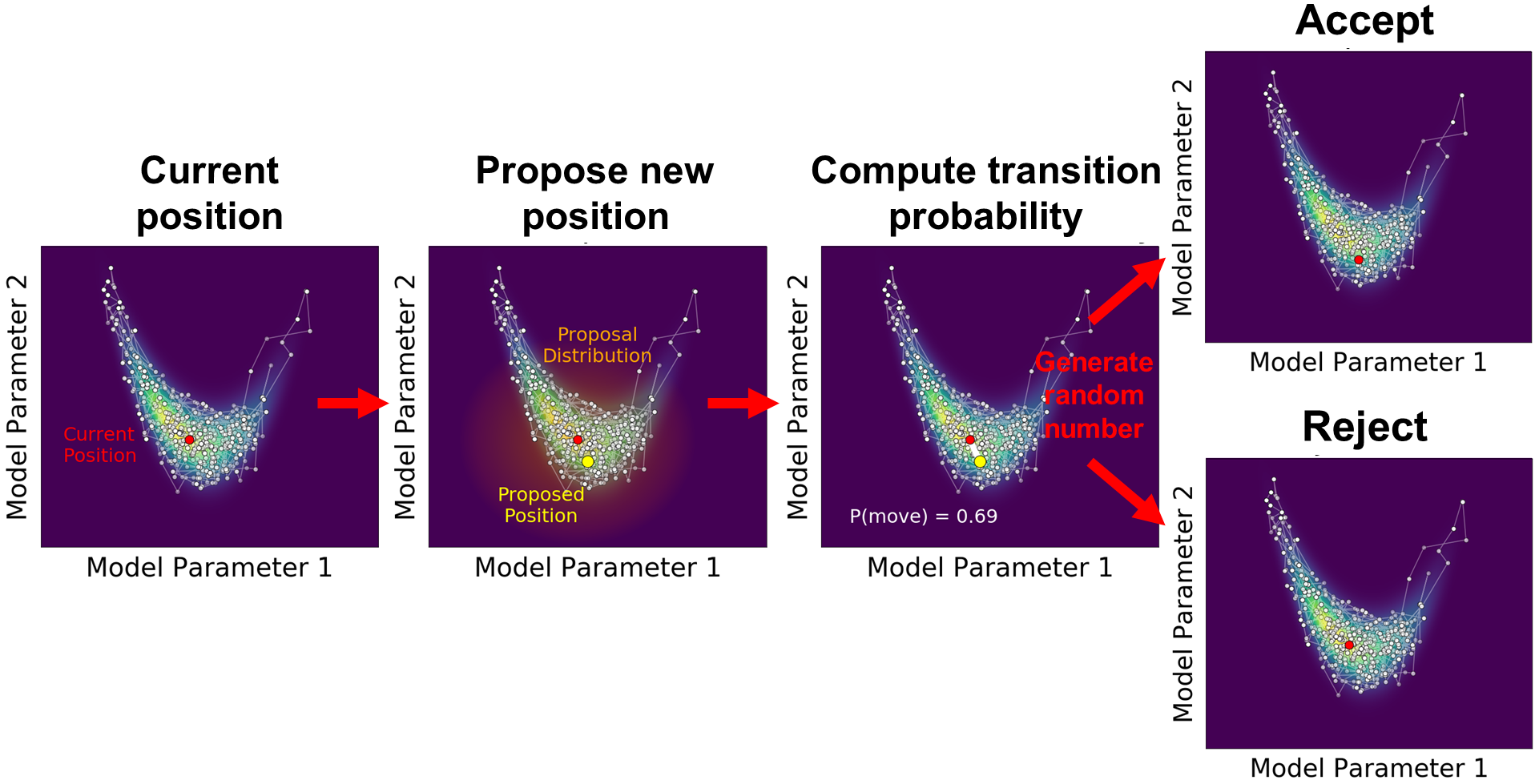}
\end{center}
\caption{A schematic illustration of the Metropolis-Hastings
algorithm. At a given iteration $i$, we have generated a chain
of samples $\{ \params_1 \rightarrow \dots \rightarrow \params_i \}$
(white) up to the current position $\params_i$ (red) whose behavior follows
the underlying posterior $\posterior(\params)$ (viridis color map).
We then propose a new position $\params_{i+1}'$ (yellow) from the proposal distribution
(orange shaded region). We then compute 
the transition probability $T(\params_{i+1}'|\params_i)$
(white) based on the posterior $\proposal(\params)$ 
and proposal $\proposal(\params'|\params)$ densities. We then
generate a random number $u_{i+1}$ uniformly from 0 to 1. If
$u_{i+1} \leq T(\params_{i+1}'|\params_i)$, we accept the move
and make our next position in the chain $\params_{i+1} = \params_{i+1}'$.
If we reject the move, then $\params_{i+1} = \params_{i}$.
See \S\ref{subsec:mh} for additional details.
}\label{fig:mh}
\end{figure}

Generating samples following this approach can be done using the
\textbf{Metropolis-Hastings (MH) Algorithm} \citep{metropolis+53_alt,hastings70}:
\begin{enumerate}
    \item \textit{Propose} a new position $\params_i \rightarrow \params_{i+1}'$
    by generating a sample from the proposal distribution
    $\proposal(\params_{i+1}'|\params_i)$.
    \item \textit{Compute} the transition probability
    $T(\params_{i+1}'|\params_i)
    = \min\left[1, \frac{\posterior(\params_{i+1}')}{\posterior(\params_i)}
    \frac{\proposal(\params_i|\params_{i+1}')}{\proposal(\params_{i+1}'|\params_i)}
    \right]$.
    \item \textit{Generate} a random number $u_{i+1}$ from $[0, 1]$.
    \item If $u_{i+1} \leq T(\params_{i+1}'|\params_i)$, \textit{accept}
    the move and set $\params_{i+1} = \params_{i+1}'$. 
    If $u_{i+1} > T(\params_{i+1}'|\params_i)$, \textit{reject} the move
    and set $\params_{i+1} = \params_i$.
    \item Increment $i = i + 1$ and repeat this process.
\end{enumerate}
See {\color{red} \autoref{fig:mh}} for a schematic illustration of this process.

Because algorithms like the MH algorithm generate a \textit{chain} of states where the
next proposed position only depends on the current position rather than
any of its past positions (i.e. it ``forgets'' the past),
they are known as \textbf{Markov processes}.
Combining these two terms with the Monte Carlo nature of simulating new
positions is what gives Markov Chain Monte Carlo (MCMC) its namesake.

An issue with generating a chain of samples in practice is the
fact that our chain only has finite length and a starting position $\params_0$. 
If our chain were infinitely long, we would expect it
to visit every possible position in parameter space, rendering
the exact starting position is unimportant. However, since in practice we terminate 
sampling after only $n$ iterations, starting from a location $\params_0$ that has
an extremely low probability means an inordinate fraction
of our $n$ samples will occupy this low-probability region, possibly biasing
our final results. Since we have limited knowledge beforehand about where
$\params_0$ is relative to our posterior, 
in practice we generally want to remove the initial
chain of states once we are confident our chain has begun sampling from
higher-probability regions. Discussing various approaches
for identifying and removing samples from this \textbf{burn-in period} 
is beyond the scope of this article; for additional information,
please see \citet{gelmanrubin92}, \citet{gelman+13}, 
and \citet{vehtari+19} along with references therein.

\subsection{Effective Sample Size and Auto-Correlation Time} \label{subsec:autocorr}

At this point, MCMC seems like it should be the optimal method for
any situation: by simulating samples directly from the (unknown) posterior,
we can achieve an optimal estimate for any expectation values we wish
to evaluate. In practice, however, this does not hold true.
MCMC values rely on specific algorithmic procedures such as 
the MH algorithm to generate samples,
whose limiting behavior \textit{reduces to} a chain of samples
$\{ \params_1 \rightarrow \dots \rightarrow \params_n \}$
whose distribution follows the posterior. Any given sample $\params_i$,
however, is more likely than not to be 
\textbf{correlated} with both the previous sample in the 
sequence $\params_{i-1}$ and the subsequent sample in the sequence
$\params_{i+1}$.

This occurs for two reasons. First, new positions $\params_i$ drawn
from $\proposal(\params_i|\params_{i-1})$ by construction tend to depend
on the current position $\params_{i-1}$. This means that the position
we propose at iteration $i+1$ from will 
be correlated with the position at iteration $i$, 
which itself will be correlated with the position at
iteration $i-1$, etc. 

Second, even if we set
$\proposal(\params'|\params)=\proposal(\params')$ so that all
of our proposed positions are uncorrelated, 
our transition probability $T(\params'|\params)$ still ensures 
that we will eventually reject the new position 
so that $\params_{i+1}=\params_{i}$.
Since samples at exactly the same position are maximally correlated,
this ensures that samples from our chain will ``on average'' have non-zero
correlations. Note that having low \textbf{acceptance fractions} 
(i.e. the fraction of proposals that are accepted rather than rejected)
will lead to a larger fraction of the chain containing 
these perfectly correlated samples, increasing the overall correlation.

As mentioned in \S\ref{subsec:ess}, correlated samples
provide less information about the underlying distribution they
are sampled from since their behavior doesn't just depend on the
underlying distribution but also the neighboring samples in the
sequence. Samples that are more highly correlated then should 
lead to a reduced ESS.

This intuition can be quantified by introducing
the \textbf{auto-covariance} $C(t)$ 
for some integer lag $t$. Assuming that we have an infinitely long chain
$\{ \params_{1} \rightarrow \dots \}$,
the auto-covariance $C(t)$ is:
\begin{equation}
    C(t) \equiv \meanwrt{(\params_{i} - \bar{\params})
    \cdot (\params_{i+t} - \bar{\params})}{i}
    = \lim_{n\rightarrow\infty} \frac{1}{n}
    \sum_{i=1}^{n} (\params_{i} - \bar{\params}) \cdot (\params_{i+t} - \bar{\params})
\end{equation}
where $\cdot$ is the dot product. In other words, 
we want to know the covariance between $\params_i$ at some iteration
$i$ and $\params_{i+t}$ at some other iteration $i+t$, averaged
over all all possible pairs of samples $(\params_i, \params_{i+t})$
in our infinitely long chain. Note that the amplitude $|C(t)|$
will be maximized at $|C(t=0)|$,
where the two samples being compared are identical,
and minimized with $|C(t)|=0$ when 
$\params_{i}$ and $\params_{i+t}$ are completely independent
from each other.

Using the auto-covariance, we can define the corresponding
\textbf{auto-correlation} $A(t)$ as
\begin{equation}
    A(t) \equiv \frac{C(t)}{C(0)}
\end{equation}
This now measures the average degree of correlation between 
samples separated by an integer lag $t$. In the case where $t=0$,
both samples are identical and $A(t=0) = 1$. In the case
where the samples are uncorrelated over lag $t$,
$A(t) = 0$.

\begin{figure}
\begin{center}
\includegraphics[width=\textwidth]{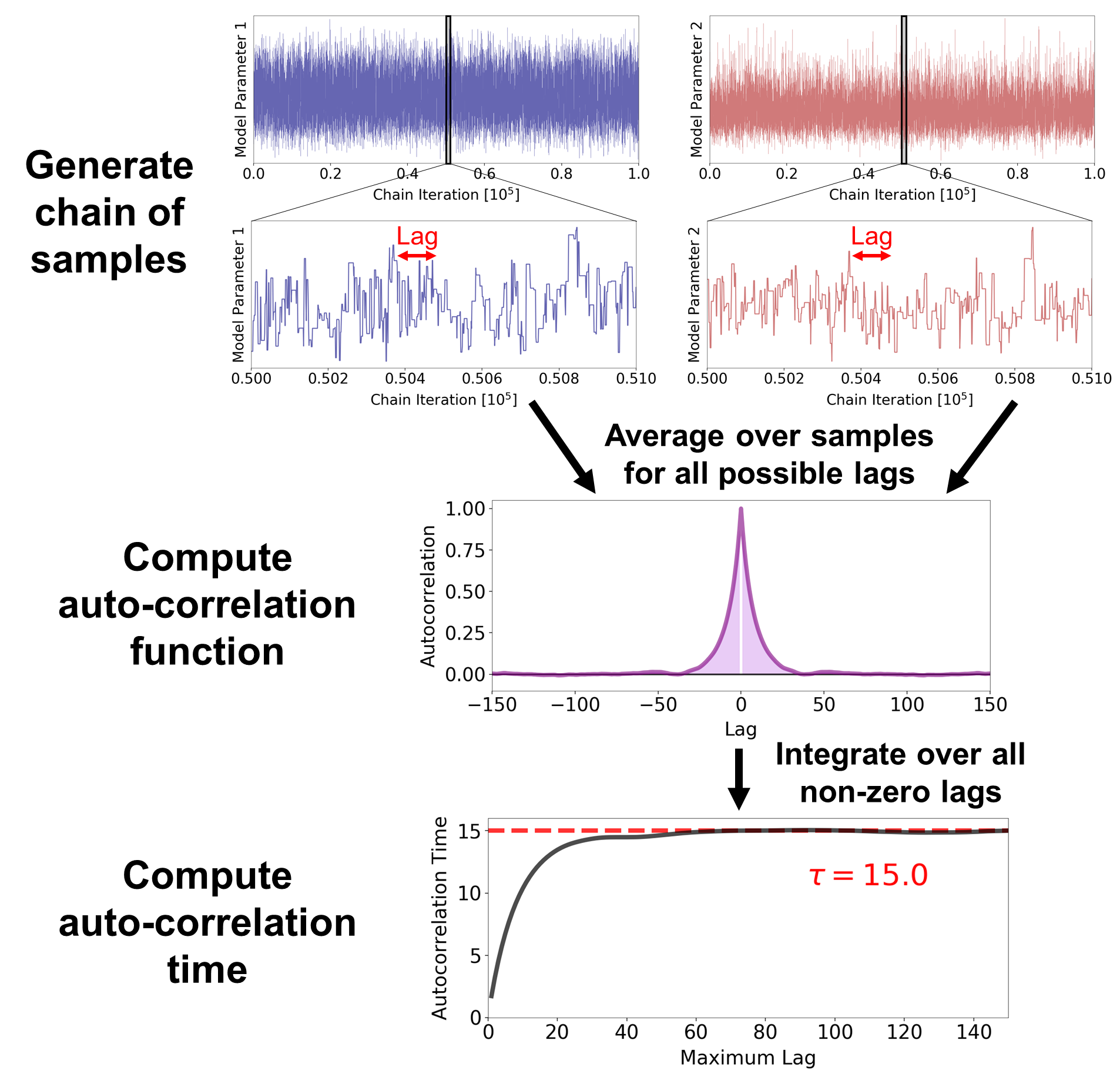}
\end{center}
\caption{A schematic illustration of the auto-correlation
associated with MCMC. MCMC methods generate a chain of
samples $\{ \params_1 \rightarrow \dots \rightarrow \params_n \}$ (top),
but these tend to be strongly correlated on small length scales (top middle).
We can quantify the degree of correlation by
computing the corresponding auto-correlation $A(t)$
over our set of samples and all possible time lags $t$
(bottom middle). This quantity is $1$ when $t=0$
and drops to $0$ as $t \rightarrow \pm \infty$. The overall
auto-correlation time $\tau$ associated with our chain of
samples is then just the integrated auto-correlation over $t \neq 0$.
See \S\ref{subsec:autocorr} for additional details.
}\label{fig:autocorr}
\end{figure}

The overall \textbf{auto-correlation time}
for our chain is just the auto-correlation $A(t)$ summed over
all non-zero lags ($t \neq 0$):
\begin{equation}
    \tau \equiv \sum_{t=-\infty}^{\infty} A(t) - 1
    = 2 \sum_{t=1}^{\infty} A(t)
\end{equation}
where the $-1$ comes from the fact that
the auto-correlation with no lag is just $A(t=0) = 1$
(i.e. each sample perfectly correlates with itself) and
the substitution arises from the fact that $A(t) = A(-t)$ by symmetry.
If $\tau = 0$, then it takes
no time at all for samples to become uncorrelated and the
samples can be assumed to be iid. If $\tau > 0$, then
it takes on average $\tau$ additional iterations for
samples to become uncorrelated. 
An illustration of this process is shown in 
{\color{red} \autoref{fig:autocorr}}.

Incorporating the auto-correlation time leads directly
to a modified definition for the ESS:
\begin{equation}
    n_{\rm eff}' \equiv \frac{n_{\rm eff}}{1 + \tau}
\end{equation}
In practice, we cannot precisely compute $\tau$ since
we do not have an infinite number of samples and do not know
$\posterior(\params)$. Therefore we often need to
generate an estimate $\hat{\tau}$ of the auto-correlation time
using the existing set of $n$ samples we have. While discussing
various approaches taken to derive $\hat{\tau}$ is beyond the
scope of this work, please see \citet{brooks+11} 
for additional details.

The fact that MCMC methods are subject to non-negative
auto-correlation times ($\tau \geq 0$) but have optimal importance
weights $\tilde{w}_i = 1$ give an ESS of
\begin{equation}
    n_{\rm eff,MCMC}' = \frac{n_{\rm eff,MCMC}}{1 + \tau}
    = \frac{n}{1 + \tau} \leq n
\end{equation}
This means that \textit{there is no guarantee
that MCMC is always the optimal choice to achieve
the largest ESS}. In particular,
Importance Sampling methods, which can generate
fully iid samples with no auto-correlation time ($\tau = 0$)
but non-optimal importance weights $\tilde{w}_i$, instead have
an ESS of
\begin{equation}
    n_{\rm eff,IS}' = \frac{n_{\rm eff,IS}}{1 + \tau}
    = n_{\rm eff,IS} 
    = \frac{\left(\sum_{i=1}^{n} \tilde{w}_i\right)^2}
    {\sum_{i=1}^{n} \tilde{w}_i^2} \leq n
\end{equation}
which can be greater than $n_{\rm eff, MCMC}'$ at fixed $n$.

Given the results above, it should now be clear
that \textit{the central motivating concern of MCMC methods
is whether they can generate a chain of samples
with an auto-correlation time small enough to outperform
Importance Sampling.} Whether or not
this is true will depend on the posterior, the approach used to generate
the chain of samples (see \S\ref{subsec:mh} and \S\ref{sec:example})
and the proposal distribution $\proposal(\params)$ 
used for Importance Sampling (see \S\ref{subsection:samp_strat}).

\subsection*{Exercise: MCMC over a 2-D Gaussian} \label{exercise:mcmc}

\subsubsection*{Setup}

Let's again return to our examples from 
\S\ref{sec:grid} and \S\ref{sec:montecarlo}, in which
our unnormalized posterior is well-approximated by a 2-D Gaussian (Normal)
distribution:
\begin{equation*}
    \tilde{\posterior}(x,y) 
    = \exp\left\{-\frac{1}{2}\left[\frac{(x-\mu_x)^2}{\sigma_x^2}
    + \frac{(y-\mu_y)^2}{\sigma_y^2}\right]\right\}
\end{equation*}
where $(\mu_x,\mu_y)=(-0.3,0.8)$ and $(\sigma_x^2,\sigma_y^2)=(2,0.5)$.

We want to use MCMC to approximate various posterior
integrals from this distribution.
We will start by choosing our proposal distribution $\proposal(x',y'|x,y)$
to be a 2-D Gaussian with a mean of $0$ and standard deviation of $1$:
\begin{equation*}
    \proposal(x',y'|x,y) = \Normal{(\mu_x,\mu_y)=(x,y)}{(\sigma_x,\sigma_y)=(1,1)}
\end{equation*}

\subsubsection*{Parameter Estimation}

Using the above proposal, generate $n=1000$ samples following the MH algorithm starting
from the position $(x_0,y_0)=(0,0)$. Using these samples,
compute an estimate of the means $\meanwrt{x}{\posterior}$
and $\meanwrt{y}{\posterior}$ as well as the corresponding 68\% credible
intervals (or closest approximation)
$[x_{\rm low}, x_{\rm high}]$ and $[y_{\rm low}, y_{\rm high}]$.
How accurate are each of these quantities compared with the values we might
expect?

\subsubsection*{Evidence Estimation}

Next, use a set of $10 \times 10$ bins from
$x=[-5, 5]$ and $y=[-5, 5]$ to construct an estimate $\rho(x,y)$ 
from the resulting set of samples. Using this estimate
for the density, compute an estimate of the evidence $\evidence$.
How accurate is our approximation? Does it substantially
change if we adjust the number and/or size of the bins?

\subsubsection*{Auto-Correlation Time and Effective Sample Size}

Use numerical methods to compute an estimate of the
auto-correlation time $\tau$ and the corresponding effective sample size $n_{\rm eff}$.
How efficient is our sampling ($n_{\rm eff}/n$) compared to the default
Importance Sampling approach from the exercise in \S\ref{sec:montecarlo}?
Does this mirror what we'd expect given the acceptance fraction of our proposals?
What do these quantities this tell us about how well 
our proposal $\proposal(x,y)$ matches the structure of the underlying posterior
$\posterior(x,y)$?

\subsubsection*{Uncertainties}

Repeat the above exercises $m=30$ times to get an
estimate for how much our estimates of each quantity can vary.
Is the variation in line with what might be expected given 
the typical effective sample size?

\subsubsection*{Consistency and Convergence}

Now repeat the above exercise using $n=2500$ and $n=10000$ samples
points and comment on any differences.
How much has the overall accuracy improved? Do
the estimates appear convergent and consistent as $n_{\rm eff}$ increases? 
How much do the errors on quantities shrink as a function 
of $n$ and/or $n_{\rm eff}$? Is this similar or different
from the observed dependence from the Importance Sampling exercise
in \S\ref{sec:montecarlo}?

\subsubsection*{Sampling Efficiency}

Next, adjust the $(\sigma_x, \sigma_y)$ of the proposal distribution
to try and improve $n_{\rm eff}$ at fixed $n$. How close is the final
ratio $\sigma_x/\sigma_y$ of our proposal to that of the underlying
posterior? Are there any additional scaling differences between the rough
size of our proposal $\proposal(x',y'|x,y)$ relative to the
underlying posterior $\posterior(x,y)$?
Given that $\tilde{\posterior}(x,y)$ may differ from the
structure assumed when picking $\proposal(x',y'|x,y)$,
can you think of any possible scheme to try and adjust our proposal
using an existing set of samples?

\subsubsection*{Burn-In}

Finally, adjust the starting position to be at
$(x_0,y_0)=(10,10)$ instead of $(0,0)$ and generate a new chain of 
samples. Plot the $x$ and $y$ positions of the chain over time.
Are there any obvious signs of the burn-in period? How many samples
roughly should be assigned to burn-in and subsequently removed from
our chain? Are there any possible heuristics that might help to identify
the initial burn-in period?

\section{Sampling the Posterior with MCMC} \label{sec:sampling}

The approach by which MCMC methods are able to generate a chain of
samples immediately gives a mental image of our chain ``exploring'' the
posterior. While it is true that the density of samples from the chain
$\rho(\params) \rightarrow \posterior(\params)$ as $n \rightarrow \infty$, 
\textit{the primary purpose of MCMC is estimating expectation values}
$\meanwrt{f(\params)}{\posterior}$.
Although this might seem like a subtle difference,
this distinction is actually crucial for understanding how MCMC
algorithms (should) behave in practice. We discuss this in more
detail below.

\subsection{Approximating the Posterior} \label{subsec:post_approx}

Although algorithms such as MH (\S\ref{subsec:mh}) are constructed to
ensure the density of the chain of samples $\rho(\params)$
generated by MCMC converges to the posterior $\posterior(\params)$
as $n \rightarrow \infty$, this \textit{does not} necessarily translate
into an efficient method to approximate the posterior in practice. In
other words, $n$ might need to be extremely large for this constraint to hold.
So how many samples do we need to ensure $\rho(\params)$ is a good approximation
to $\posterior(\params)$?

To start, we first need to define some metric for what a ``good''
approximation is. A reasonable one might be that we would like
to know the posterior within some region $\delta_{\params}$
to within some precision $\epsilon$ so that
\begin{equation}
    \left| \frac{1}{n} \sum_{i=1}^{n} \indicator{\params_i \in \delta_{\params}} 
    - \int_{\delta_{\params}} \posterior(\params) \deriv \params \right|
    \equiv |\hat{p}(\delta_{\params}) - p(\delta_{\params})| < \epsilon
\end{equation}
where $p(\delta_{\params})$ is the total probability contained within
$\delta_{\params}$ and $\hat{p}(\delta_{\params})$ is the fraction of
the MCMC chain of samples contained within the same region. While it might
seem strange to only estimate this for one region, I will shortly
generalize this to encompass the
entire\footnote{Technically the procedure outlined in this section
only works for finite volumes. The basic intuition, however, holds even
when parameters are unbounded although proving those results is beyond
the scope of this work.} posterior.

In the ideal case where our samples are iid and drawn from $\posterior(\params)$,
our samples each have a probability $p(\delta_{\params})$ 
of being within $\delta_{\params}$.
The probability that $\hat{p}(\delta_{\params}) = m/n$ then follows the
\textbf{binomial distribution}:
\begin{equation}
    P\left(\hat{p}(\delta_{\params}) = \frac{m}{n} \right) 
    = \binom{n}{m} \left[p(\delta_{\params})\right]^m
    \left[1 - p(\delta_{\params}) \right]^{n-m}
\end{equation}
In other words, our samples end up inside $\delta_{\params}$ a total of $m$ times
with probability $p(\delta_{\params})$ and outside $\delta_{\params}$ a
total of $n-m$ times with probability $1 - p(\delta_{\params})$. The
additional binomial coefficient $\binom{n}{m}$ for ``$n$ choose $m$'' 
accounts for all possible unique cases where $m$ samples can
end up within $\delta_{\params}$ out of our total sample size of $n$.

This distribution has a mean of $p(\delta_{\params})$, so for
any finite $n$ we expect $\hat{p}(\delta_{\params})$ to be an
\textbf{unbiased estimator} of $p(\delta_{\params})$:
\begin{equation}
    \mean{\hat{p}(\delta_{\params}) - p(\delta_{\params})} 
    = p(\delta_{\params}) - p(\delta_{\params}) = 0
\end{equation}
The variance, however, depends on the sample size:
\begin{equation}
    \mean{|\hat{p}(\delta_{\params}) - p(\delta_{\params})|^2}
    = \frac{p(\delta_{\params}) \left[1 - p(\delta_{\params})\right]}{n}
\end{equation}

In practice, we can expect there to be some non-zero auto-correlation
time $\tau > 0$. This will increase the number of MCMC samples
we will need to generate to be confident that our estimate
$\hat{p}(\delta_{\params})$ is well-behaved. Inserting a factor
of $1+\tau$ and substituting our expectation value from above
into our accuracy constraint then gives a
rough constraint for the number of samples $n$ we would require
as a function of $\epsilon$:
\begin{equation}
    n \gtrsim 
    \frac{p(\delta_{\params}) \left[1 - p(\delta_{\params})\right]}
    {\epsilon^2/(1+\tau)} 
    \sim \frac{\hat{p}(\delta_{\params}) 
    \left[1 - \hat{p}(\delta_{\params})\right]}
    {\epsilon^2} \times (1+\hat{\tau})
\end{equation}
The final substitution of $p(\delta_{\params})$ and $\tau$ with their
noisy estimates $\hat{p}(\delta_{\params})$ and $\hat{\tau}$
arises from the fact that
in practice we don't know $p(\delta_{\params})$ or 
$\tau$ (both of which require full knowledge of the posterior).
We are therefore forced to rely on estimators
derived from our set of $n$ samples.

Let's now examine this result more closely. As expected,
the total number of samples is proportional
to $1 + \hat{\tau}$: if it takes longer to generate independent
samples, then we need more samples to be confident we have characterized
the posterior well in a given region. 
We also see that $n \propto \epsilon^{-2}$, so that
if we want to reduce the error by a factor of $x$ we need
to increase our sample size by a factor of $x^2$.

The behavior in the numerator is more interesting. Note
that $\hat{p}(\delta_{\params}) \left[1 - \hat{p}(\delta_{\params})\right]$
is maximized for $\hat{p}(\delta_{\params}) = 0.5$, and so the largest
sample size needed is when we have split our posterior directly in half.
In all other cases the sample size needed will be smaller because
there will be more samples outside or inside the region of interest
whose information we can leverage. The exact value of $\hat{p}(\delta_{\params})$
of course depends on both the posterior $\posterior(\params)$
and the target region $\delta_{\params}$: the sample size needed to approximate
the posterior to some $\epsilon$ 
near the peak of the distribution (the small region where $\posterior(\params)$ is large)
will likely be different than
the sample size needed to accurately estimate the tails of the
distribution (the large region where $\posterior(\params)$ is small).

While the above argument holds if we are looking to estimate the
posterior in just \textit{one} region, ``converging to the posterior''
implies that we want $\rho(\params)$
to become a good approximation to $\posterior(\params)$ \textit{everywhere}.
We can enforce this new requirement by splitting our posterior into $m$
different sub-regions $\{ \delta_{\params_1}, \dots, \delta_{\params_m} \}$ 
and requiring that each sub-region is well constrained:
\begin{equation}
    |\hat{p}(\delta_{\params_1}) - p(\delta_{\params_1})| < \epsilon_1
    \quad\quad \dots \quad\quad
    |\hat{p}(\delta_{\params_m}) - p(\delta_{\params_m})| < \epsilon_m
\end{equation}
Substituting in the expected errors on each of these
constraints then gives us an approximate limit
on the number of samples $n_j$ that we need
to estimate the posterior in each region $\delta_{\params_j}$:
\begin{equation}
    n_j \gtrsim \frac{\hat{p}(\delta_{\params_j}) 
    \left[1 - \hat{p}(\delta_{\params_j})\right]}
    {\epsilon_j^2} \times (1+\hat{\tau})
\end{equation}
The total number of samples we need is then simply:
\begin{equation}
    n \gtrsim 
    \sum_{j=1}^{m} n_j
\end{equation}

This approach of dividing up our posterior into sub-regions
is conceptually similar to the grid-based approaches
described in \S\ref{sec:grid}. As such,
it is also subject to the same drawbacks:
we expect the number of regions $m$ to increase
\textit{exponentially} with the number of dimensions $d$.
For instance, if we just wanted to divide our posterior up
into $m$ \textbf{orthants} we would end up with $m=2^d$ regions:
2 in 1-D (left-right), 4 in 2-D (upper-left, lower-left,
upper-right, lower-right), 8 in 3-D, etc.

This effect implies that we should in general expect
the number of samples required to ensure $\rho(\params)$ is a good
approximation to $\posterior(\params)$ for some specified accuracy $\epsilon$
to scale as
\begin{equation}
    n \gtrsim k^d
\end{equation}
where $k$ is a constant that depends on the accuracy requirements.
This puts approximating the full posterior firmly in the 
``curse of dimensionality'' regime (see \S\ref{subsec:curse}).\footnote{
A direct corollary of this result is that, while the
evidence estimates from MCMC \textit{are} consistent, 
the rate of convergence to the underlying value 
will proceed exponentially more slowly
as $d$ increases.}

While many practitioners talk about MCMC being 
an efficient method to ``approximate the posterior'', in practice it is rarely
used to approximate $\posterior(\params)$ directly.
As discussed in \S\ref{sec:what} and shown in
{\color{red} \autoref{fig:corner}}, almost all quantities 
that are reported in the literature \textit{do not} rely on approximations to
the full $d$-dimensional posterior, but rather approximations to marginalized
distributions that are almost always restricted to no more 
than $k\lesssim3$ parameters at a time.
The act of marginalizing over the remaining $d-k$ parameters
helps to counteract the curse of dimensionality illustrated here.
While it is technically fair to say that MCMC can ``explore'' 
the marginalized $k$-D posteriors for certain limited sets of parameters, 
this type of language can often lead to more misconceptions than insights.

\subsection{Posterior Volume} \label{subsec:volume}

The basic consequences outlined in \S\ref{subsec:post_approx} 
are more general than the specific case where we imagine dividing
up the posterior into orthants or other regions.
Fundamentally, computing any expectation
over the posterior $\meanwrt{f(\params)}{\posterior}$ requires
integrating over the \textit{entire domain} of our parameters 
$\params$. We therefore want to understand how the \textbf{volume}
of this domain behaves (i.e. how many parameter combinations there are).
Once we have a grasp on how this behaves, we can then starting trying
to quantify how this will impact our estimates.

To start, let's consider the
$d$-dimensional hyper-cube (the $d$-cube)
with side length $\ell$ in all $d$ dimensions. Its volume scales as
\begin{equation}
    V(\ell) = \prod_{i=1}^{d} \ell = \ell^d
\end{equation}
The differential volume element between $\ell$ and
$\ell + \deriv \ell$ is
\begin{equation}
    \deriv V(\ell) = (d \times \ell^{d-1}) \times (\deriv\ell) \propto \ell^{d-1}
\end{equation}

This exponential scaling with dimensionality means that volume becomes
increasingly concentrated in thin shells located in regions located
progressively further away from the center of the $d$-cube. 
As an example, consider the length-scale
\begin{equation}
    \ell_{50} = 2^{-1/d} \ell
\end{equation}
that divides the $d$-cube into two equal-sized regions
with 50\% of the volume contained interior to $\ell_{50}$ and
50\% of the volume exterior to $\ell_{50}$. 
In 1-D, this gives $\ell_{50}/\ell = 0.5$ as we'd
expect. In 2-D, this gives $\ell_{50}/\ell \approx 0.7$. In 3-D,
$\ell_{50}/\ell \approx 0.8$. In 7-D, $\ell_{50}/\ell \approx 0.9$.
By the time we get to 15-D, we have $\ell_{50} /\ell\approx 0.95$,
which means that 50\% of the volume is located in the last 5\% of the 
length-scale near the boundary of the $d$-cube.
While the constants may change when considering other shapes (e.g., spheres),
in general this exponential scaling as a function of $d$ is a generic feature
of higher-dimensional volumes. In other words, increasing the number
of parameters leads to an exponential increase in the number of
available parameter combinations that we have to explore.

\begin{figure}
\begin{center}
\includegraphics[width=\textwidth]{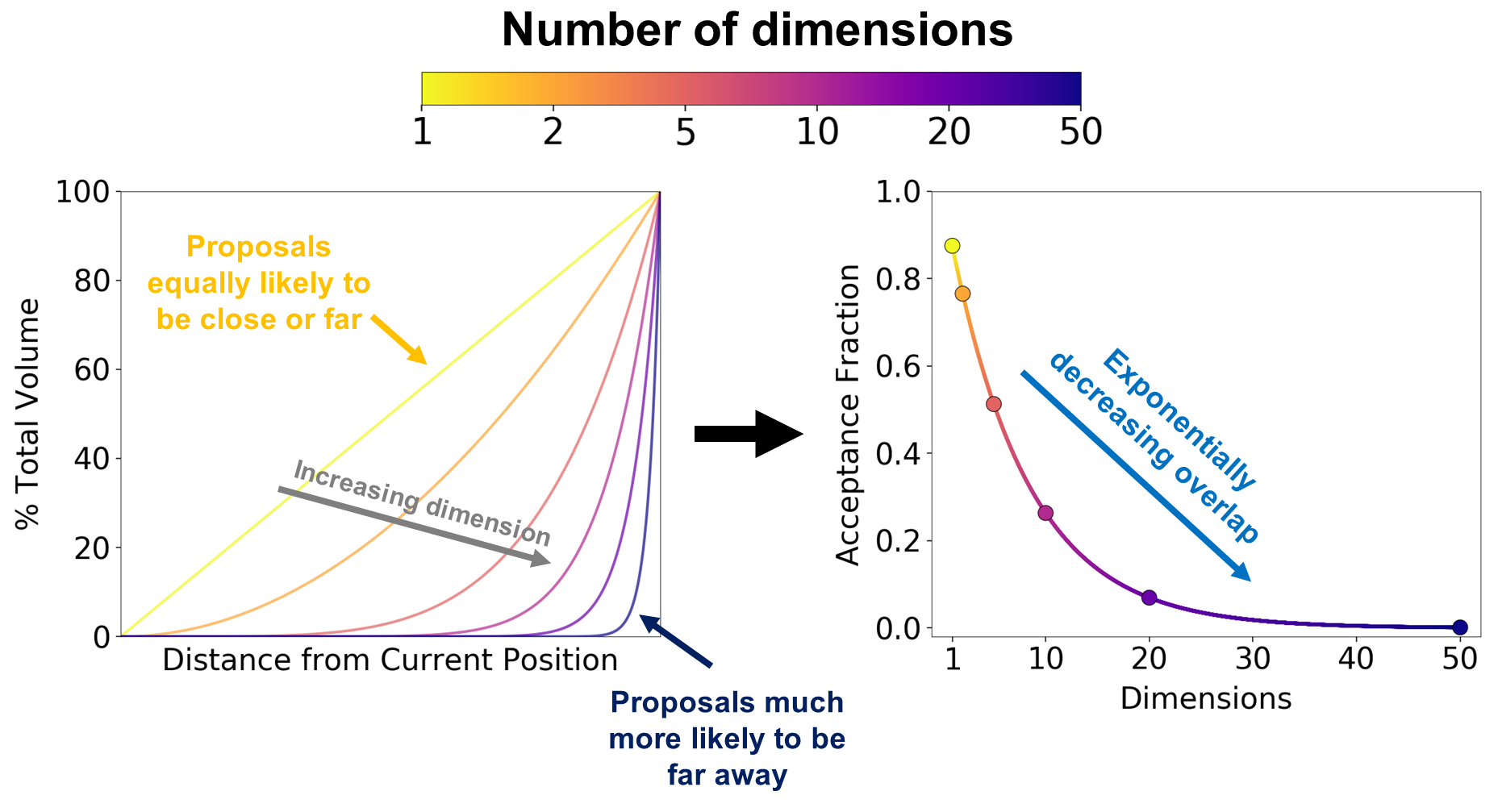}
\end{center}
\caption{A schematic illustration of how the curse of dimensionality
affects MCMC acceptance fractions via posterior volume.
At a given position position $\params$, 
the volume increases $\propto r^d$ as a function 
of distance $r$ away from that position (left). 
As the dimensionality increases, this implies volume
becomes concentrated progressively further out, leading to 
larger distances between proposed positions $\params'$ and the current
position $\params$. Most of these positions have significantly lower
posterior probabilities $\posterior(\params')$ compared to the 
current value $\posterior(\params)$, leading to an
exponential decline in the typical acceptance fraction
(and a corresponding increase in the auto-correlation time)
as the dimensionality increases (right). Adjusting the size
and/or shape of the proposal $\proposal(\params'|\params)$
can help to counteract this behavior.
See \S\ref{subsec:volume} for additional details.
}\label{fig:vol}
\end{figure}

In addition to affecting the long-term behavior of MCMC, 
this exponential increase in volume also directly
impacts how MCMC methods operate. To see why this is the case, we
need look no further than the transition probability used in the
MH algorithm discussed in \S\ref{subsec:mh}:
\begin{equation*}
    T(\params_{i+1}|\params_i)
    \equiv \min\left[1, \frac{\posterior(\params_{i+1})}{\posterior(\params_i)}
    \frac{\proposal(\params_i|\params_{i+1})}{\proposal(\params_{i+1}|\params_i)}
    \right]
\end{equation*}
The non-trivial portion of this expression cleanly splits into two terms.
The first is dependent on the \textit{volume} and is related to
how we proposed our next position from $\proposal(\params'|\params)$.
The second is dependent on the \textit{density} and is related to
how the posterior density changes between the two positions.

In practice, our transition probability can be interpreted as
a basic corrective approach: after proposing
a new position from some nearby volume, we then try to ``correct''
for differences between our proposal and the underlying posterior
by only accepting these moves sometimes based on changes in the underlying density.
In high dimensions, this basic ``tug of war'' between the volume (proposal) and
the density (posterior) can break down as the vast majority of an object's 
volume becomes concentrated near the outer edges.\footnote{Alternative
methods such as Hamiltonian Monte Carlo \citep{neal12} can get around
this problem by smoothly incorporating changes in the density and volume.}
For instance, in the case where our
proposal $\proposal(\params'|\params)$ is a cube with side-length $\ell$
centered on $\params$, this leads to a median length-scale of
$\ell_{50} = 2^{-1/d} \ell$, which increases rapidly from $0.5 \ell$
to $\approx \ell$ as the dimensionality increases. The same logic
also applies to other proposal distributions (see \S\ref{sec:example}).
This focus on positions either far away or with very similar separation
length-scales as $\ell_{50} \rightarrow \ell$ 
means that many choices of $\proposal(\params'|\params)$
have a tendency to ``overshoot'', proposing new positions with much
smaller posterior densities compared to the current position. These
new positions are then almost always rejected, leading to extremely
low acceptance fractions and correspondingly
long auto-correlation times. An example of this effect is
illustrated in {\color{red} \autoref{fig:vol}.}

One of the main ways to counteract this behavior is to
adjust the size/shape of the proposal $\proposal(\params'|\params)$
so that the fraction of proposed positions that are accepted remains
sufficiently high. This helps to ensure the posterior density 
$\posterior(\params)$ does not change too drastically when proposing 
positions new positions, leading to lower overall auto-correlation times.
Details of how to implement these schemes in practice are beyond 
the scope of this article; please see
\textbf{citation} for additional details.

\subsection{Posterior Mass and Typical Sets} \label{subsec:mass}

Above, I described how the behavior of volume in high dimensions
can impact the performance of our MCMC MH sampling algorithm,
possibly leading to inefficient proposals and 
low acceptance fractions. Let's assume that we have resolved this
problem and have an efficient way of generating our chain
of samples. We now have a secondary question: \textit{where are these
samples located?}

From our discussion in \S\ref{subsec:post_approx}, we know that
the highest \textit{density} of samples $\rho(\params)$
will be located where the posterior density $\posterior(\params)$
is also correspondingly high. However, this region $\delta_{\params}$
might only correspond to a small portion of the posterior. 
Indeed, given there is exponentially more volume as the dimensionality
increases, it is almost guaranteed that models with many parameters
$\params$ will have the vast majority of the posterior located
outside the region of highest density. 

A consequence of this is that
the majority of samples in our chain will be located
away from the peak density. As a result, \textit{our chain
spends the majority of its time
generating samples in these regions}.
This has a huge impact in the way our chain is expected to behave:
while the highest \textit{concentration}
of samples will be located in the regions of highest
posterior density, the largest \textit{amount} of samples
will actually be located in the regions of highest \textbf{posterior
mass} (i.e. density times volume). 
Since this implies that a ``typical'' sample 
(picked at random) will most likely be located in this
region of high posterior mass,
this region is also commonly referred to
as the \textbf{typical set}.

To make this argument a little easier to conceptualize,
let's imagine that we have a 3-parameter model $\params = (x, y, z)$ 
and $\posterior(x,y,z)$ is spherically symmetric.
While we could imagine trying to
integrate over $\posterior(x,y,z)$ directly in terms of
$\deriv x \deriv y \deriv z$, it is almost always
easier to instead integrate over such a distribution
in ``shells'' with differential volume
$\deriv V(r) = 4 \pi r^2 \deriv r$
as a function of radius
$r = \sqrt{x^2 + y^2 + z^2}$. This allows us
to rewrite the 3-D integral over $(x,y,z)$ as a 1-D integral over $r$:
\begin{equation}
    \int \posterior(x,y,z) \deriv x \deriv y \deriv z
    = \int \posterior(r) 4 \pi r^2 \deriv r
    \equiv \int \posterior'(r) \deriv r
\end{equation}
where $\posterior'(r) \equiv 4 \pi r^2 \posterior(r)$ is now
the 1-D density as a function of $r$. This ``boosts'' the contribution
as a function of $r$ by the differential volume element of the
shell associated with $\posterior(r)$, and implies that the
the posterior should have some sort of shell-like structure (i.e.
$\posterior'(r)$ is maximized for $r > 0$).

Although not all posterior densities can be expected to
be spherically-symmetric in this way, in general we can
rewrite the $d$-D integral over $\params$ as a 1-D volume integral
over $V$ defined by some unknown iso-posterior 
contours\footnote{Indeed, alternative Monte Carlo methods such as 
Nested Sampling \citep{skilling04,skilling06} or 
Bridge/Path Sampling \citep{gelmanmeng98} actually are designed to 
evaluate this type of volume integral explicitly.}
\begin{equation}
    \int \posterior(\params) \deriv \params
    = \int \posterior(V) \deriv V
\end{equation}
As outlined in \S\ref{subsec:volume}, we generically
expect the size of each volume element to go as
$\deriv V \sim r^{d-1} \deriv r$ where $r$ is the
distance from the peak of posterior. So the basic
intuition we get from the simple spherically-symmetric case
still applies and we expect
\begin{equation}
    \int \posterior(V) \deriv V 
    \sim \int \posterior(r) r^{d-1} \deriv r
    = \int \posterior'(r) \deriv r
\end{equation}

\begin{figure}
\begin{center}
\includegraphics[width=\textwidth]{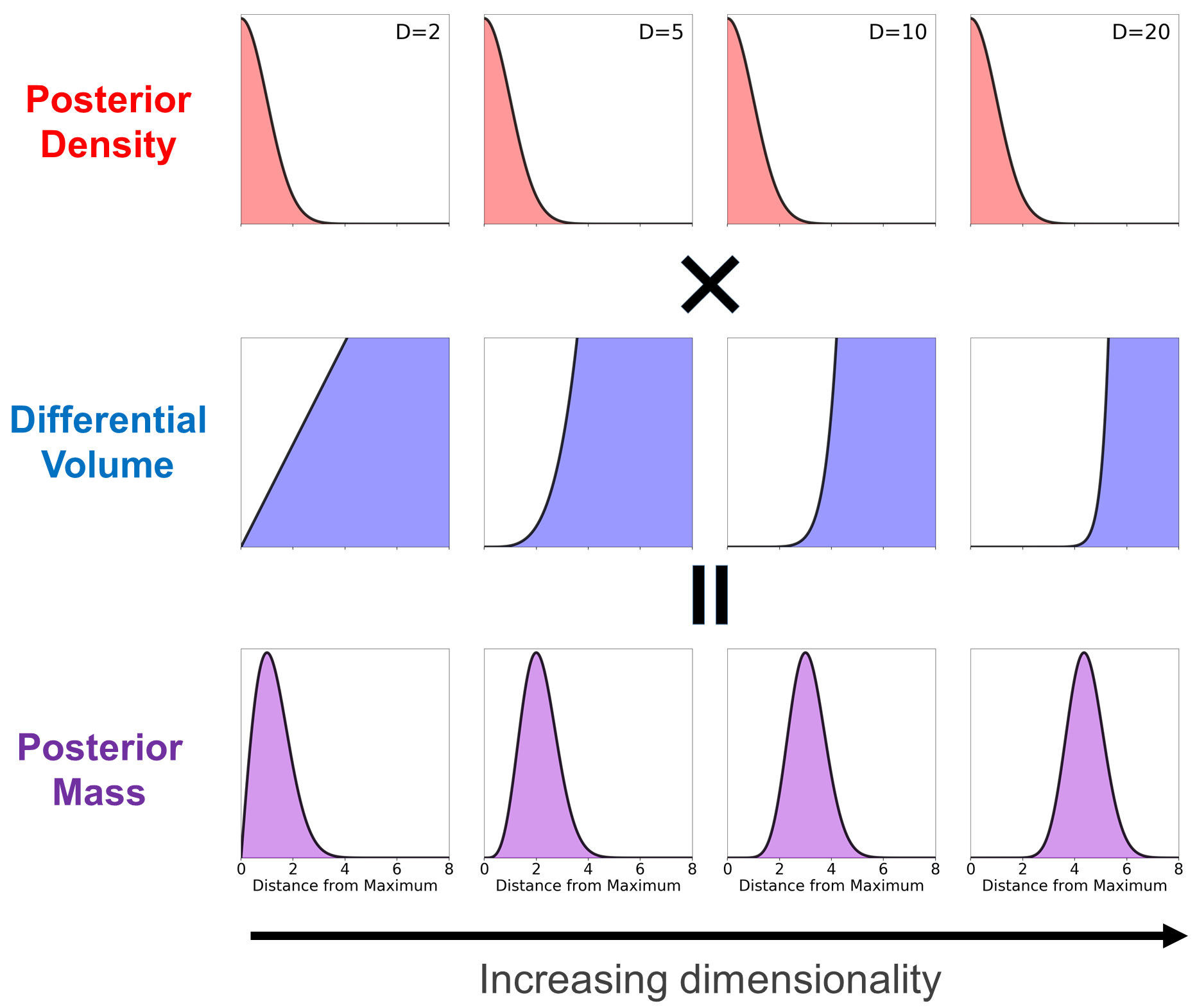}
\end{center}
\caption{A schematic illustration of how the posterior mass
behaves as a function of dimensionality using a $d$-dimensional
Gaussian. The top panel shows the posterior density 
$\posterior(r) \propto e^{-r^2/2}$ (red) plotted 
as a function of distance $r$ from the maximum posterior density at $r=0$
as the number of dimensions $d$ increases (left to right). As expected,
this distribution remains constant. The middle panel shows
the differential volume element 
$\deriv V(r) \propto r^{d-1} \deriv r$ (blue) of
the corresponding shell at radius $r$. This illustrates the
exponentially increasing volume contributed by shells further
away from the maximum. The bottom panel shows corresponding
``posterior mass'' as a function of radius 
$\posterior'(r) \propto r^{d-1} \posterior(r) \propto r^{d-1} e^{-r^2/2}$ (purple).
Due to the increasing amount of volume located further away from
the maximum posterior density, we see that the majority of the
posterior mass (and therefore of any samples we generate with MCMC)
are actually located a shell located far away from the $r=0$.
See \S\ref{subsec:mass} for additional details.
}\label{fig:mass}
\end{figure}

As before, the differential volume element of the
shell associated with $\posterior(r)$ ``boosts'' its overall
contribution as a function of $r$. This boost also becomes
exponentially stronger as $d$ increase. 
\textit{For even moderately-sized $d$,
we therefore expect the posterior mass
to be mostly contained in a thin shell located at a radius $r'$
with some width $\Delta r'$}.
See {\color{red} \autoref{fig:mass}} for an illustration of this effect
based on the toy problem presented in \S\ref{subsec:analytic}.

This result has two immediate implications. First, \textit{the majority
of our samples are not located where the posterior
density is maximized}. This is the result of an exponentially
increasing number of parameter combinations, which allow a
small handful of excellent fits to the data to
be easily overwhelmed by a substantially larger number
of mediocre fits. MCMC methods are therefore generally
inefficient at locating and/or characterizing the region of
peak posterior density.

Second, as $d$ increases we generally would expect the radius of the
shell containing the bulk of the posterior mass to increase, moving further
and further away from the peak density due to the exponentially increasing
available volume. Since the majority of our samples are located in this
region, \textit{our chain will spend the vast majority 
of time generating samples from this shell}.

This allows us to now outline exactly why it is challenging
to propose samples efficiently in high dimensions:
\begin{enumerate}
    \item To make sure our acceptance fractions remain reasonable, 
    we need to ensure our proposed positions mostly lie within 
    this shell of posterior mass.
    \item However, obtaining an independent sample requires 
    being able to (in theory) propose any position within this shell.
    \item This means that our auto-correlation time
    will principally be set by how long it takes to ``wander around''
    the shell, which will be a function of its overall size $r'$,
    its width $\Delta r'$, and the number of dimensions $d$.
\end{enumerate}

\section{Application to a Simple Toy Problem} \label{sec:example}

I now consider a concrete, detailed example
to illustrate how all the concepts discussed in \S\ref{sec:mcmc} and
\S\ref{sec:sampling} come together in practice. 
Throughout this section, I will outline a number of analytic
results and utilize several different MCMC sampling strategies
to generate chains of samples. I strongly encourage interested readers
to implement their own versions of the methods outlined here,
which can be used to reproduce the numerical results from this section
in their entirety.

\subsection{Toy Problem} \label{subsec:analytic}

In this toy problem, we will take our (unnormalized) 
posterior to be a $d$-dimensional 
Gaussian (Normal) distribution with a mean
of $\mu = 0$ and a standard deviation of $\sigma$
in all dimensions:
\begin{equation}
    \tilde{\posterior}(\params) 
    = \exp\left[-\frac{1}{2}\frac{|\params|^2}{\sigma^2}\right]
\end{equation}
where $|\params|^2 = \sum_{i=1}^{d} \Theta_i^2$ is the squared magnitude
of the position vector.

Based on the results from \S\ref{subsec:mass}, 
we can better understand the properties of this distribution by
rewriting the posterior density
in terms of the ``radius''
$r \equiv | \params | = \sqrt{\sum_{i=1}^{d} \Theta_i^2}$
away from the center:
\begin{equation}
    \tilde{\posterior}(r) = \exp\left[-\frac{r^2}{2\sigma^2}\right]
\end{equation}
The corresponding volume contained within a given radius $r$ is then
\begin{equation}
    V(r) \propto r^d
\end{equation}
The corresponding posterior mass is $\tilde{\posterior}'(r)$
is then defined via
\begin{align}
    \tilde{\posterior}(V) \deriv V(r)
    \propto e^{-r^2/2\sigma^2} r^{d-1} \deriv r \nonumber
    \equiv \tilde{\posterior}'(r) \deriv r
\end{align}
Note that this is closely related
to the \textbf{chi-square distribution}.

The \textbf{typical radius} $r_{\rm peak}$ 
where the posterior mass peaks (i.e. is maximized)
and a sample is most likely to be located
can be derived by setting $\deriv \tilde{\posterior}'(r)/\deriv r = 0$.
Solving this gives
\begin{equation}
    r_{\rm peak} = \sqrt{d-1} \sigma
\end{equation}
In other words, while in 1-D a typical sample
is most likely to be located at the peak of the 
distribution with $r_{\rm peak} = 0$, in higher dimensions
this changes quite drastically. While $r_{\rm peak} = 1\sigma$
in 2-D, it is $2\sigma$ in 5-D, $3\sigma$ in 10-D, and $5\sigma$
in 26-D. This is a direct consequence of the
huge amount of volume at larger radii in high dimensions:
although a sample at $r=5\sigma$ has a posterior density $\posterior(r)$
orders of magnitude worse than a sample at $r=0$, the enormous number of
parameter combinations (volume) available at $r=5\sigma$
more than makes up for it.

In general, we expect the posterior mass to comprise
a \textbf{``Gaussian shell''} centered at some radius
\begin{equation}
    r_{\rm mean} \equiv \meanwrt{r}{\posterior'}
    = \int_0^\infty r \posterior'(r) \deriv r
    = \sqrt{2} \frac{\Gamma\left(\frac{d+1}{2}\right)}
    {\Gamma\left(\frac{d}{2}\right)} \sigma
    \approx \sqrt{d} \sigma
\end{equation}
with a standard deviation of
\begin{equation}
    \Delta r_{\rm mean}
    \equiv \sqrt{\meanwrt{(r - r_{\rm mean})^2}{\posterior'}}
    = \sigma \sqrt{d - 2 
    \left(\frac{\Gamma\left(\frac{d+1}{2}\right)}
    {\Gamma\left(\frac{d}{2}\right)}\right)^2}
    \approx \frac{\sigma}{\sqrt{2}}
\end{equation}
where $\Gamma(d)$ is the Gamma function and the
approximations are taken for large $d$.
See {\color{red} \autoref{fig:mass}} for an
illustration of this behavior.

\subsection{MCMC with Gaussian Proposals} \label{subsec:mcmc_gauss}

Let us now consider a chain of samples
$\{ \params_1 \rightarrow \dots \rightarrow \params_n \}$.
The distance between two samples
$\params_m$ and $\params_{m+t}$ 
separated by some lag $t$ will be
\begin{equation}
    |\params - \params'|
    = \sqrt{\sum_{i=1}^{d} (\Theta_{m,i} - \Theta_{m+t,i})^2}
\end{equation}
Assuming that the lag $t \gg \tau$ is
substantially larger than the auto-correlation time $\tau$,
we can assume each sample is approximately iid distributed following
our Gaussian posterior. This then gives an expected separation of
\begin{equation}
    \Delta r_{\rm sep} 
    \equiv \sqrt{\meanwrt{|\params_{m} - \params_{m+t}|^2}{\posterior}}
    = \sqrt{\sum_{i=1}^{d} \meanwrt{(\Theta_{m,i} - \Theta_{m+t,i})^2}{\posterior}}
    = \sqrt{2d} \sigma \approx \sqrt{2} r_{\rm mean}
\end{equation}

We can in theory propose samples in such a way so that the separation
$|\params_{i+1} - \params_{i}|$
between a proposed position $\params_{i+1}$ 
and the current position $\params_i$ follows the
ideal separation of $\sqrt{2}r_{\rm mean}$ derived above
by using a simple Gaussian proposal distribution:
\begin{equation}
    \proposal(\params_{i+1}|\params_i)
    \propto \exp\left[-\frac{1}{2}\frac{|\params_{i+1} - \params_i|^2}{2\sigma^2}\right]
\end{equation}
While this proposal has the same \textit{shape} as the posterior,
it is centered on $\params_i$ rather than $0$. Using our intuition for
how volume behaves based on \S\ref{subsec:volume}, we can
conclude that the majority of samples proposed from 
this choice of $\proposal(\params'|\params)$ will probably have little overlap
with the posterior. 

\begin{figure}
\begin{center}
\includegraphics[width=\textwidth]{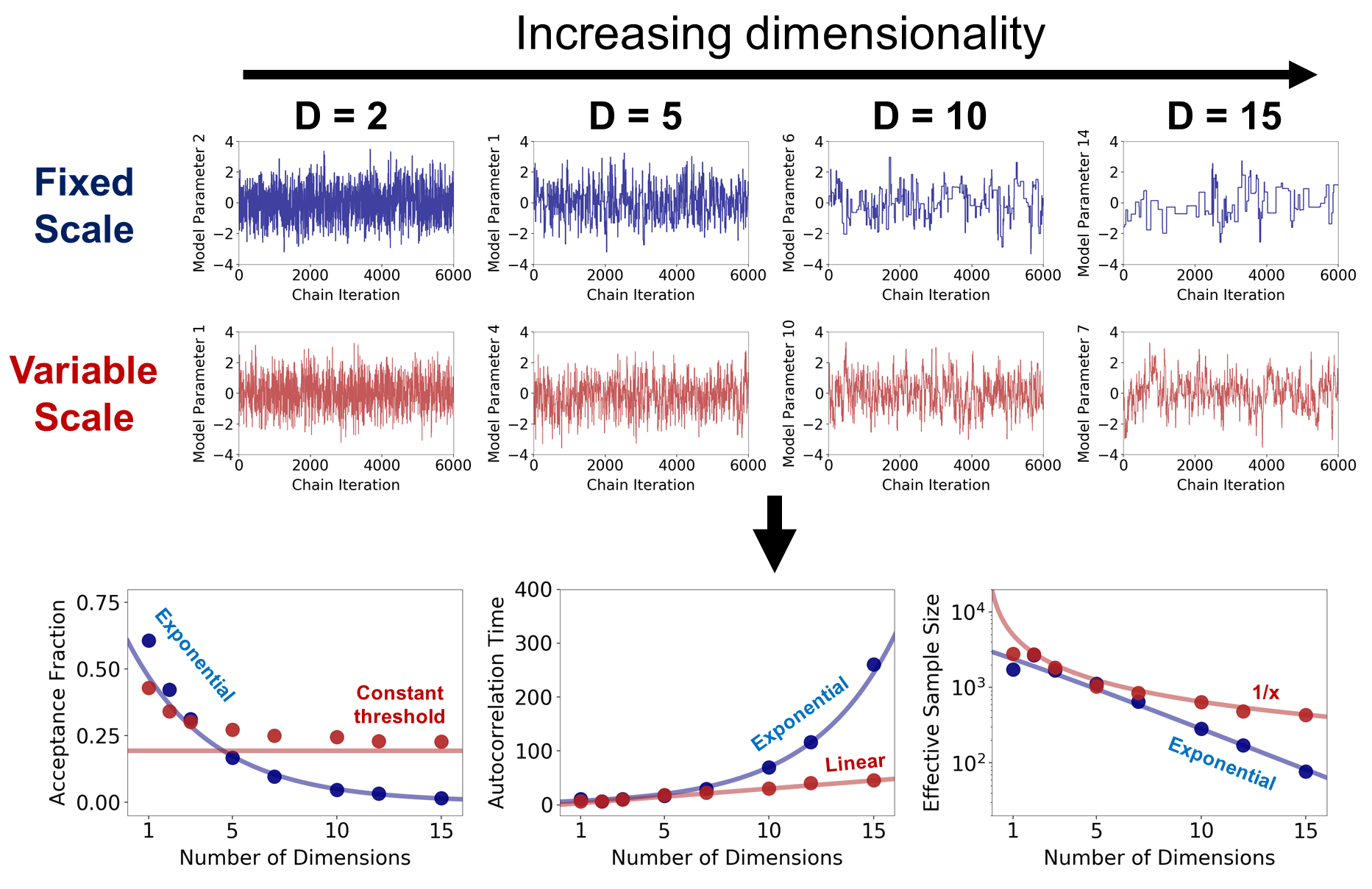}
\end{center}
\caption{Numerical results showcasing the performance
of a simple MH MCMC sampler with Gaussian proposals
on our toy problem, a $d$-dimensional Gaussian with mean $\mu=0$
and standard deviation $\sigma=1$ in every dimension.
The top series of panels show snapshots of a random parameter
from the chain as a function of dimensionality (increasing
from left to right) assuming an unchanging proposal with
constant scale factor $\gamma=\sqrt{2}$ (blue) 
and a shrinking proposal with $\gamma=2.5/\sqrt{d}$ designed
to target a constant acceptance fraction of $\sim 25\%$ (red).
The bottom panels show the corresponding
acceptance fractions (left), auto-correlation times (middle), and
effective sample sizes (right) from our chains (colored points)
as a function of dimensionality. The approximations from 
\S\ref{subsec:mcmc_gauss} are shown as light colored lines.
Shrinking the size of the proposal helps to keep samples 
within the bulk of the posterior mass, substantially reducing
the auto-correlation time and increasing the effective sample size.
Failing to do so leads to
an exponentially decreasing fraction of good
proposals and a corresponding exponential increase/decrease in the
auto-correlation time/effective sample size.
See \S\ref{subsec:mcmc_gauss} for additional discussion.
}\label{fig:mcmc_gauss}
\end{figure}

Indeed, numerical simulation suggests
the typical fraction of positions that will be accepted given the above proposal
roughly scales as
\begin{equation}
    \langle f_{\rm acc}(d) \rangle
    \equiv \exp\left[\meanwrt{\ln T(\params_{i+1}|\params_i)}
    {\posterior,\proposal}\right]
    \sim \exp\left[-\frac{d}{4} - \frac{1}{2}\right]
\end{equation}
which decreases exponentially as the dimensionality increases,
similar to {\color{red} \autoref{fig:vol}}. Likewise, we find
the auto-correlation time roughly scales as
\begin{equation}
    \langle \tau(d) \rangle 
    \equiv \exp\left[\meanwrt{\ln \tau}
    {\posterior,\proposal}\right]
    \sim \exp\left[\frac{d}{4} + \frac{7}{4}\right]
\end{equation}
This exponential dependence
arises because the overlap between the typical Gaussian proposal
$\proposal(\params'|\params)$ and the underlying posterior 
$\posterior(\params)$ essentially reduces to the small volume
where two thin shells overlap. Since the radii of the shells
goes as $\propto \sqrt{d}$ while the widths remain roughly constant,
the ``fractional size'' of the shell (and the corresponding overlap)
ends up decreasing exponentially.

To counteract this effect, we need to adjust the $\sigma$ of our
proposal distribution by some factor $\gamma$:
\begin{equation}
    \proposal_\gamma(\params_{i+1}|\params_i)
    \propto \exp\left[-\frac{1}{2}\frac{|\params_{i+1} - \params_i|^2}
    {(\gamma\sigma)^2}\right]
\end{equation}
where our previous proposal assumes $\gamma = \sqrt{2}$.
If we want to ensure our typical acceptance fraction will remain
roughly constant as a function of dimension $d$, $\gamma$ needs to scale as
\begin{equation}
    \langle f_{\rm acc}(\gamma(d)) \rangle \approx C
    \quad \Rightarrow \quad
    \gamma(d) \propto \frac{1}{\sqrt{d}}
\end{equation}
which inversely tracks the expected radius $r_{\rm mean}$
of the typical set. We find that taking
\begin{equation}
    \gamma = \frac{\delta}{\sqrt{d}}
\end{equation}
leads to a typical acceptance fraction of
\begin{equation}
    \langle f_{\rm acc}(\delta/\sqrt{d}) \rangle
    \approx \exp\left[-\left(\frac{\delta^2}{4}\right)^{2} 
    - \frac{\delta}{2}\right]
\end{equation}
as $d$ becomes large with a typical auto-correlation time of
\begin{equation}
    \langle \tau(\delta/\sqrt{d}) \rangle \approx 3 d
\end{equation}
for reasonable choices of $\delta$.
This linear dependence is a substantial improvement over
our earlier exponential scaling.

\subsubsection*{Numerical Tests} \label{subsubsec:sims_1}

To confirm these results, I sample from this $d$-dimensional
Gaussian posterior (assuming $\sigma=1$ for simplicity) using
two MH MCMC algorithms for $n=20,000$ iterations
based on these proposal distributions.
The first proposes new points 
assuming $\gamma=\sqrt{2}$. The second
assumes $\gamma=2.5/\sqrt{d}$ in order
to maintain a roughly constant acceptance fraction of 25\%. 
As shown in {\color{red} \autoref{fig:mcmc_gauss}}, the chains
behave as expected given our theoretical predictions as a function
of dimensionality, with the constant proposal quickly becoming stuck
while the adaptive proposal continues sampling normally. While
the auto-correlation time $\tau$ increases in both cases, 
the increase in the latter case (where it is driven by decreasing
size/scale of the proposal distribution)
is much more manageable than the former (where it is driven by
the exponentially decreasing acceptance fraction).

\subsection{MCMC with Ensemble Proposals} \label{subsec:mcmc_ensemble}

One drawback to the Gaussian proposals explored above is that
we have to specify the structure of the distribution ahead of time.
In this specific case, we assumed that:
\begin{enumerate}
    \item the width of the posterior in each dimension (parameter)
    was constant such that $\sigma_1 = \sigma_2 = \dots = \sigma_n = \sigma$ and
    \item the parameters were entirely uncorrelated with each other such that
    the correlation coefficient $\rho_{ij} = 0$ between any two dimensions
    $i$ and $j$.
\end{enumerate}

In general, there is no good reason to assume that either of these are true.
This means we have to also estimate the entire set of
$d(d+1)/2$ free parameters that determine the overall covariance
structure of our unknown posterior distribution.
Trying to adjust the covariance structure in order
to improve our sampling efficiency and decrease the auto-correlation
time (see \S\ref{exercise:importance} and \S\ref{exercise:mcmc})
becomes one of the most difficult parts of running MCMC algorithms in practice.

While there are schemes to perform these adjustments during
an extended burn-in period (see, e.g., \citealt{brooks+11}), there
is significant appeal in methods that can ``auto-tune'' without
much additional input from the user. One class of such approaches
are known as \textbf{ensemble} or \textbf{particle} methods.
These methods attempt to use many $m$ chains running
simultaneously (i.e. in parallel) to improve the performance 
of any individual chain.

\begin{figure}
\begin{center}
\includegraphics[width=\textwidth]{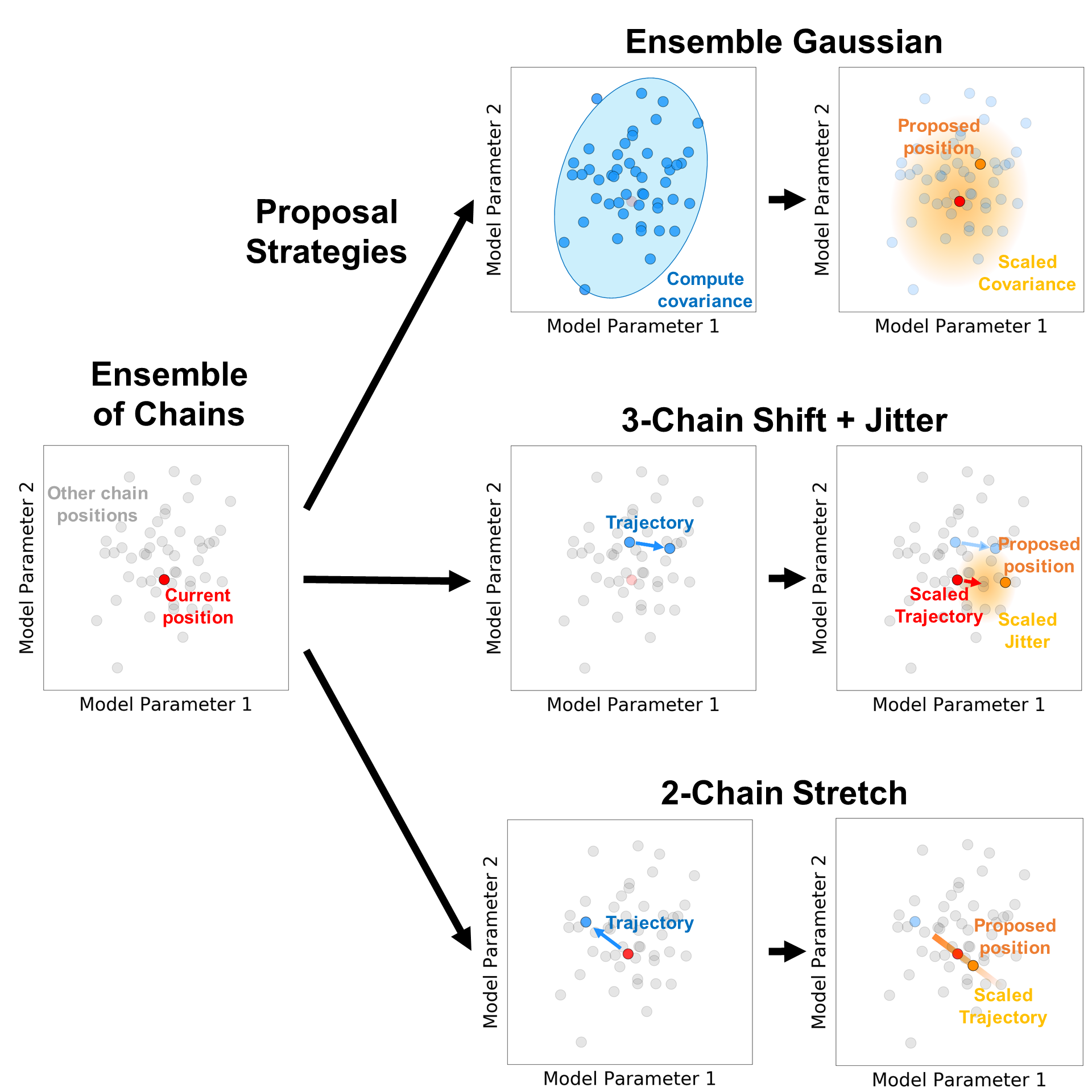}
\end{center}
\caption{A schematic illustration of the three ensemble MCMC
methods described in \S\ref{subsec:mcmc_ensemble}. 
The current state of the chain we are interested in updating (red)
and the other chains in the ensemble (gray) are shown on the left.
In the top panels (ensemble Gaussian; \S\ref{subsubsec:ensemble_gauss}),
we compute the covariance of the other $k \neq j$
chains (middle) and use a scaled version to subsequently
propose a new position. 
In the middle panels (3-chain shift + jitter; \S\ref{subsubsec:demcmc}),
we use two additional chains $k \neq l \neq j$ to compute a trajectory. 
We then propose a new position based on this scaled trajectory plus
a small amount of ``jitter''. 
In the bottom panels (2-chain stretch; \S\ref{subsubsec:emcee}), we use
only one additional chain $k \neq j$ to propose a new trajectory. We
then propose a random position along a scaled version of this
trajectory with the proposal probability 
varying as a function of scale.
See \S\ref{subsec:mcmc_ensemble} for additional details.
}\label{fig:mcmc_particles}
\end{figure}

We explore three variations of ensemble methods here
that attempt to exploit $m \gtrsim d(d+1)/2$ chains running
simultaneously:
\begin{enumerate}
    \item using the ensemble of particles to condition
    a Gaussian proposal distribution,
    \item using trajectories from multiple particles along with
    Gaussian ``jitter'', and
    \item using affine-invariant transformations of
    trajectories from multiple particles.
\end{enumerate}
A schematic illustration of these 
methods is shown in {\color{red} \autoref{fig:mcmc_particles}}.

As we might expect, an immediate drawback of these methods is
they rely on having enough particles to characterize
the overall structure of the space (i.e. the curse of dimensionality).
While this limits their utility when sampling from high-dimensional spaces, 
they can be attractive options in moderate-dimensional spaces 
($d \lesssim 25$) where a few hundred particles
are often sufficient to ensure reasonable performance.

\subsubsection{Gaussian Proposal} \label{subsubsec:ensemble_gauss}

The first approach is simply a modified
Gaussian proposal: at any iteration $i$ for any chain
$j$, we propose a new position $\params_{i+1}^{j}$
based on the current position $\params_{i}^{j}$
using a Gaussian proposal
\begin{equation}
    \proposal_\gamma^j(\params_{i+1}^{j}|\params_{i}^j)
    \propto \exp\left[-\frac{1}{2}
    (\params_{i+1}^{j} - \params_{i}^{j})^{\rm T}
    (\gamma^2\cov_i^j)^{-1}
    (\params_{i+1}^{j} - \params_{i}^{j})\right]
\end{equation}
where ${\rm T}$ is the transpose operator and
\begin{equation}
    \cov_i^j = {\rm Cov}\left[\{\params_i^1, \dots, 
    \params_i^{j-1}, \params_i^{j+1}, \dots, \params_i^m\}\right]
\end{equation}
is the empirical covariance matrix estimated from the
current positions of the $m$ chains \textit{excluding} chain $j$.
We repeat this process for each of the $m$ chains in turn.

In other words, at each iteration $i$ we want to update all
$m$ chains. We do so by updating each chain $j$ in turn based on what
the other chains are currently doing. Assuming the current position of
each chain is distributed following the underlying posterior
$\posterior(\params)$, it is straightforward to show that
$\cov_i^j$ is a reasonable approximation to the unknown
covariance structure of our posterior. In addition,
because we exclude $j$ when computing $\cov_i^j$, this proposal
is symmetric going from $\params_{i}^{j} \rightarrow \params_{i+1}^{j}$
and from $\params_{i+1}^{j} \rightarrow \params_{i}^{j}$.
This means that we satisfy detailed balance and do not
have to incorporate any proposal-dependent factors when
computing the transition probability.

\subsubsection{Ensemble Trajectories with a Gaussian Proposal} 
\label{subsubsec:demcmc}

The approach taken in \S\ref{subsubsec:ensemble_gauss}
solves the problem of trying to tune
the covariance of our initial Gaussian proposal. However, it still
assumes that a Gaussian proposal is the optimal solution.
A more general approach is one that does not rely on assuming a
proposal explicitly, but rather only relies on the distribution of
the remaining particles. 

One such approach used in the literature is
\textbf{Differential Evolution} MCMC \citep[DE-MCMC;][]{stornprice97,terbraak06}.
The main idea behind DE-MCMC is to rely on the \textit{relative positions}
of the chains at a given iteration $i$ when making new proposals. We first
randomly select two other particles $k$ and $l$ where
$\params^j_i \neq \params^k_i \neq \params^l_i$. We then propose
a new position based on the vector distance between the other
two particles $\params^k_i - \params^l_i$ with some scaling $\gamma$
along with some additional ``jitter'' $\epsilon$:
\begin{equation}
    \params_{i+1}^{j} 
    = \params_i^j + \gamma \times (\params^{k}_i - \params^l_i + \epsilon)
\end{equation}

In the case where the behavior of chains $k$ and $l$ are approximately
independent of each other and assuming the
underlying posterior distribution $\posterior(\params)$
is Gaussian with some unknown
mean $\meanvec$ and covariance $\cov$
(and ``standard deviation'' $\cov^{1/2}$), it is straightforward to
show that the distribution of $\params^k_i - \params^l_i$ will then follow
\begin{equation}
    \params^k_i - \params_l \sim \Normal{\mathbf{0}}{(2\cov)^{1/2}}
\end{equation}
Typically, the jitter $\epsilon$ is chosen to also be Gaussian distributed
with covariance $\cov_\epsilon$ such that
\begin{equation}
    \epsilon \sim \Normal{\mathbf{0}}{\cov_\epsilon^{1/2}}
\end{equation}
In general, $\cov_\epsilon$ is mostly
used to try and avoid issues caused by finite particle
sampling: since the number of unique trajectories (ignoring symmetry) is
\begin{equation*}
    n_{\rm traj} 
    = \binom{m-1}{2} 
    = \frac{(m-1)!}{2!(m-3)!}
    = \frac{(m-1)(m-2)}{2}
\end{equation*}
if $m$ is sufficiently small the DE-MCMC
procedure can only explore a small number of possible trajectories
at any given time, leading to extremely inefficient sampling.

Combined, this implies that the
proposed position has a distribution of
\begin{equation}
    \params_{i+1}^{j} 
    \sim \Normal{\params_i^j}{\gamma \times (2\cov+\cov_\epsilon)^{1/2}}
\end{equation}
This shows that the 3-particle DE-MCMC procedure can generate
new positions in a manner analogous to the ensemble Gaussian proposal
we first discussed.

\subsubsection{Affine-Invariant Transformations of Ensemble Trajectories}
\label{subsubsec:emcee}

Another approach used in the literature 
\citep[e.g., {\texttt{emcee}};][]{foremanmackey+13_alt}
is the \textbf{Affine-Invariant ``stretch move''} from 
\citet{goodmanweare10}. This 
uses only one additional particle $\params_i^k$ rather than two:
\begin{equation}
    \params_{i+1}^{j} 
    = \params_i^k + \gamma \times (\params^{j}_i - \params^k_i)
\end{equation}
In place of the jitter term $\epsilon$ from DE-MCMC,
the stretch move instead injects some amount of 
randomness by allowing $\gamma$ to vary.
By sampling $\gamma$ from some probabilty distribution 
$g(\gamma)$, we allow the proposals
to explore various ``stretches'' of the direction vector.
As shown in \citet{goodmanweare10}, if this function is chosen such that
\begin{equation}
    g(\gamma^{-1}) = \gamma \times g(\gamma)
\end{equation}
then this proposal is symmetric. Typically, $g(\gamma)$ is chosen
to be
\begin{equation}
    g(\gamma|a) = 
    \begin{cases}
    \gamma^{-1/2} & a^{-1} \leq \gamma \leq a \\
    0 & {\rm otherwise}
    \end{cases}
\end{equation}
where $a=2$ is often taken as a typical value.
Note that when $\gamma = 1$, this move leaves 
$\params_{i+1}^j = \params_i^j$ unchanged.

Compared to DEMCMC, the stretch move appears to have one clear advantage:
it doesn't have any reliance on some ``jitter'' term $\epsilon$ that reintroduces
scale-dependence into the proposal. That makes the proposal invariant to
affine transformations and only
sensitive to \textit{a single parameter} $a$, which governs the range of
scales the stretch factor $\gamma$ is allowed to explore.

This lack of jitter, however, is not substantially advantageous in practice. 
As noted in \S\ref{subsubsec:demcmc},
$\epsilon$ is really designed to avoid possible
degeneracies due to the limited number of available trajectories. In that
case we had $(m-1)(m-2)/2 \sim m^2/2$ possible trajectories; here, however, we only
have $m$ (since $\params^j_i$ is always included). This is a \textit{much}
smaller number of possible trajectories at a given $m$, 
making this particular proposal more susceptible to that particular effect.

\begin{figure}
\begin{center}
\includegraphics[width=\textwidth]{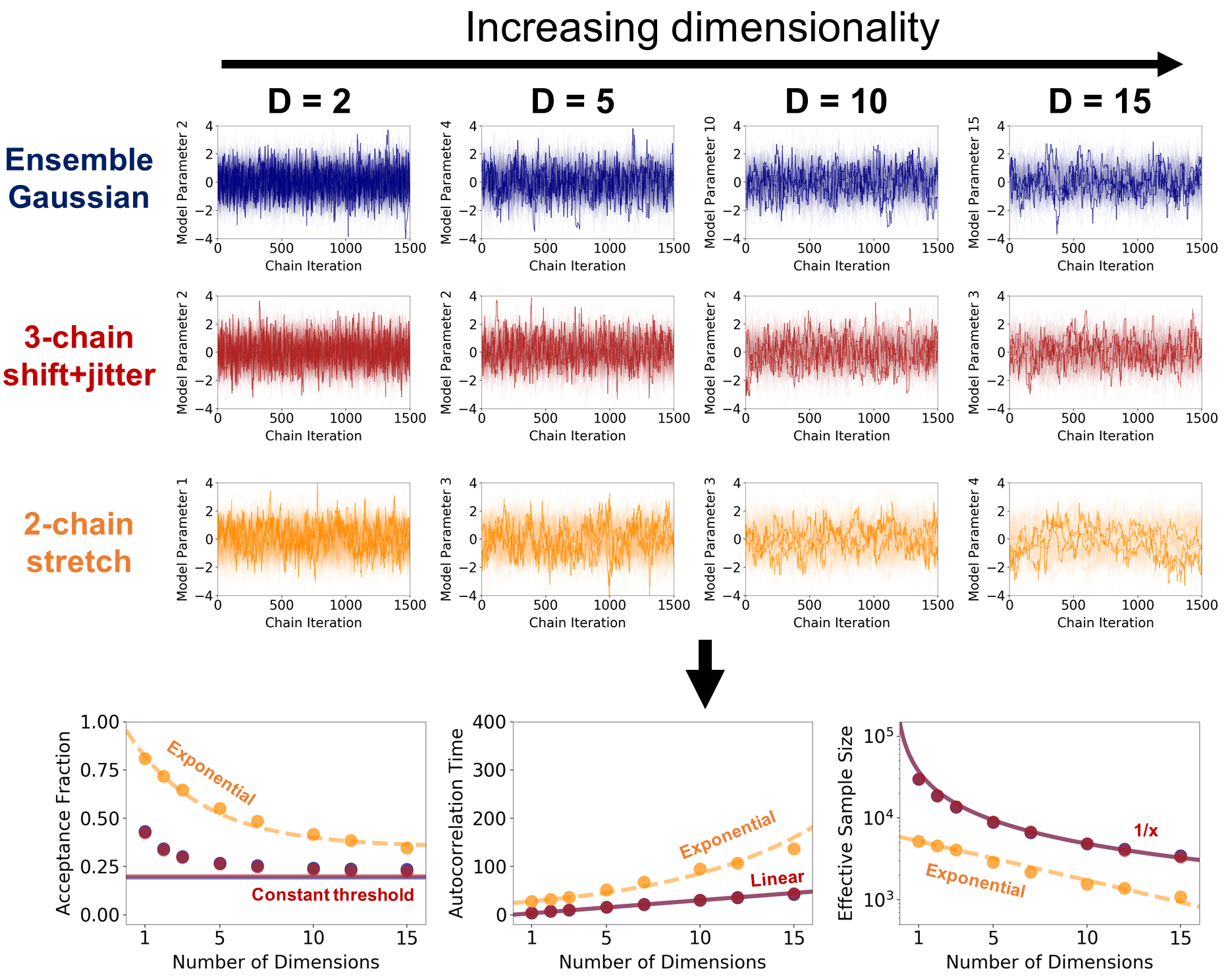}
\end{center}
\caption{Numerical results showcasing the performance
of several ensemble MH MCMC samplers on our toy problem,
a $d$-dimensional Gaussian with mean $\mu=0$
and standard deviation $\sigma=1$ in every dimension.
The top series of panels show snapshots of a random parameter
from the collection of chains (with a few chains highlighted)
as a function of dimensionality (increasing
from left to right) assuming ensemle Gaussian
proposals with $\gamma=2.5/\sqrt{d}$ (blue), 
3-chain ``shift and jitter''
proposals with $\gamma=1.7/\sqrt{d}$ (red), 
and 2-chain ``stretch'' proposals with
$\gamma$ drawn from the distribution $g(\gamma|a)$ with $a=2$
as described in \S\ref{subsubsec:emcee} (orange).
The bottom panels show the corresponding
acceptance fractions (left), auto-correlation times (middle), and
effective sample sizes (right) from our chains (colored points)
as a function of dimensionality. Approximations based on the 
\S\ref{subsec:mcmc_gauss} are shown as light solid colored lines,
with dashed lines showing rough fits.
The first two methods, which allow the
size of the proposal to shrink, are able to propose samples 
within the bulk of the posterior mass.
The last method, which is unable to do so,
instead proposes exponentially fewer good positions as the
dimensionality increases.
See \S\ref{subsec:mcmc_ensemble} for additional details.
}\label{fig:mcmc_ensemble}
\end{figure}

In addition, because this proposal involves adjusting $\gamma$ and therefore
the length of the trajectory itself, we need to consider how changing $\gamma$
affects the total \textit{volume} of the sphere centered on $\params^j_i$
with radius $\params^k_i-\params^j_i$. As discussed in \S\ref{subsec:volume},
the differential volume increases as $r^{d-1}$. Therefore, increasing
or decreasing $\gamma$ substantially adjusts the differential volume
in our proposal. This involves introducing a steep boost/penalty into our
transition probability, which now becomes:
\begin{equation}
    T(\params_{i+1}^{j}|\params_i^j, \gamma)
    = \min\left[1, \gamma^{d-1} 
    \frac{\posterior(\params^j_{i+1})}{\posterior(\params_i^j)}\right]
\end{equation}
This heavily favors proposals with $\gamma > 1$ (outwards) and heavily
disfavors proposals with $\gamma < 1$ as $d$ increases to account for the
exponentially increasing volume at larger radii.

Finally, while this stretch move actually generates proposals 
in the right overall \textit{direction}, it is not efficient at generating
samples within the bulk of the posterior mass
as the dimensionality increases.
As discussed in \S\ref{subsec:mcmc_gauss}, given the typical
position of $\params_i^j$, the typical length-scale of
the proposed positions needs to shrink by
$\propto 1/\sqrt{d}$ in order to guarantee our new sample
remains within the bulk of the posterior mass.
However, the form for $g(\gamma|a)$ specified above 
instead ensures that $\gamma$ will always be between $1/a$ and $a$.
Even if we attempt to account for this effect by letting 
$a(d) \rightarrow 1$ as $d \rightarrow \infty$ in order to
target a constant acceptance fraction and ensure more overlap, 
the asymmetry of our proposal and the $\gamma^{d-1}$ term
in the transition probability systematically biases our
proposed and accepted positions compared with the ideal distribution.
This subsequently leads to lager auto-correlation times, mostly
counteracting any expected gains.

\subsubsection*{Numerical Tests} \label{subsubsec:sims_2}

To confirm these results, I sample from this $d$-dimensional
Gaussian posterior (assuming $\sigma=1$ for simplicity) 
using each of these ensemble MH MCMC algorithms with for $n=1500$
iterations with $m=100$ chains.
In the first case, I propose a new position for chain $j$ 
at iteration $i$ using a Gaussian distribution with
a covariance $\gamma^2 \cov_i^j$ computed over the remaining ensemble of
$k \neq j$ chains, where the scale factor $\gamma=2.5/\sqrt{d}$ is
chosen to target a constant acceptance fraction of roughly 25\%. 
In the second case, I propose new positions using the DE-MCMC algorithm
with a scale factor of $\gamma=1.7/\sqrt{d}$ 
and additional Gaussian jitter with covariance $\cov_\epsilon = \cov_i^j / 5$
derived from the remaining chains in the ensemble, again targeting
an acceptance fraction of roughly 25\%.
In the third case, I propose new positions using the affine-invariant stretch
move assuming the typical form for $g(\gamma|a)$ with $a=2$.\footnote{Allowing
$a(d)$ to vary as a function of dimensionality to target a roughly constant
acceptance fraction gives similar results.}

As shown in {\color{red} \autoref{fig:mcmc_ensemble}}, the chains
behave as expected given our theoretical predictions as a function
of dimensionality. Similar to the adaptive Gaussian case,
the first two approaches continue 
sampling efficiently even as $d$ increases. The affine-invariant
stretch move, however, experiences exponentially-decreasing efficiency
and struggles to sample the posterior effectively.

\subsection{Additional Comments} \label{subsec:comments}

Before concluding, I wish to emphasize that the toy problem explored in this
section should only be interpreted as a \textit{tool} to build intuition surrounding
how certain methods are expected to behave in a controlled environment.
While the behavior as a function of dimensionality
helps to illustrate common issues, in practice 
the performance of any method will depend on the specific problem, tuning parameters,
the time spent on tuning, and many other possible factors. Since it is always possible
to find problems for which any particular method
will perform well or poorly, I encourage users to try out a variety of
approaches to find the ones that work best for their problems.

\section{Conclusion} \label{sec:conc}

Bayesian statistical methods have become increasingly prevalent in
modern scientific analysis as models have become more complex.
Exploring the inferences we can draw from these models often
requires the use of numerical techniques, the most popular of
which is known as \textbf{Markov Chain Monte Carlo (MCMC)}.

In this article, I provide a conceptual introduction
to MCMC that seeks to highlight the \textit{what}, \textit{why},
and \textit{how} of the overall approach. I first give
an overview of Bayesian inference and discuss \textit{what}
types of problems Bayesian inference generally is trying to solve, 
showing that most quantities we are interested in computing
require integrating over the posterior density.
I then outline approaches to computing these integrals using
grid-based approaches, and illustrate how adaptively changing
the resolution of the grid naturally transitions into 
the use of Monte Carlo methods. I illustrate how
different sampling strategies affect the overall efficiency
in order to motivate \textit{why} we use MCMC methods.
I then discuss various details related to \textit{how} MCMC methods 
work and examine their expected overall behavior based on
simple arguments derived from how volume and posterior density behave
as the number of parameters increases. 
Finally, I highlight the impact this conceptual understanding
has in practice by comparing the performance of various MCMC methods 
on a simple toy problem.

I hope that the material in this article, 
along with the exercises and applications,
serve as a useful resource that helps
build up intuition for how MCMC and other
Monte Carlo methods work.
This intuition should be helpful
when making decisions over when to apply MCMC methods to your own
problems over possible alternatives,
developing novel proposals and sampling strategies, 
and characterizing what issues you might
expect to encounter when doing so.

\section*{Acknowledgements}

JSS is grateful to Rebecca Bleich for continuing to tolerate his
(over-)enthusiasm for sampling during their time together. 
He would also like to thank a number of people for
helping to provide much-needed feedback during earlier stages
of this work, including Catherine Zucker, Dom Pesce, Greg Green,
Kaisey Mandel, Joel Leja, David Hogg, Theron Carmichael, and Jane Huang.
He would also like to thank Ana Bonaca, Charlie Conroy, Ben Cook,
Daniel Eisenstein, Doug Finkbeiner, Boryana Hadzhiyska, 
Will Handley, Locke Patton, and Ioana Zelko
for helpful conversations surrounding the material.

JSS also wishes to thank Kaisey Mandel and
the Institute of Astronomy at the University
of Cambridge, Hans-Walter Rix and
the Galaxies and Cosmology Department at
the Max Planck Institute for Astronomy, and Ren\'{e}e
Hlo\u{z}ek, Bryan Gaensler, and the Dunlap Institute
for Astronomy and Astrophysics at the University of Toronto
for their kindness and hospitality while hosting him over
the period where a portion of this work was being completed.

JSS acknowledges financial support from the National Science Foundation
Graduate Research Fellowship Program (Grant No. 1650114)
and the Harvard Data Science Initiative.

\bibliography{ref}
\bibliographystyle{aasjournal}



\end{document}